\documentclass{llncs}

\RequirePackage{color}
\usepackage{enumerate}
\usepackage{amssymb}
\usepackage{amsmath}
\usepackage{epsfig}
\usepackage{pst-tree}
\usepackage{pst-node}
\usepackage{pst-tree}
\usepackage{subfigure}
\usepackage{alltt}
\usepackage{multicol}

\newcommand{\tab}[1]{\ifvmode\else\\\fi\hspace*{#1em}}

\def\punto{$\hspace*{\fill}\Box$}
\newcommand{\nop}[1]{}
\newcommand{\tuple}[1]{{\langle#1\rangle}}

\def\tv{t}
\def\tid{d}

\spnewtheorem*{proviso}{Proviso}{\itshape}{\rmfamily}

\def\punto{$\hspace*{\fill}\Box$}

\title{World-set Decompositions:\\ Expressiveness and Efficient
  Algorithms\thanks{This article is an extended version of the paper
\cite{AKO06iWSD} that has appeared in the Proceedings
of the International Conference
on Database Theory (ICDT) 2007.}}

\author{Dan Olteanu\inst{1}, Christoph Koch\inst{2}, and Lyublena Antova\inst{2}}
\institute{%
Oxford University Computing Laboratory\\dan.olteanu@comlab.ox.ac.uk
\and
Department of Computer Science, Cornell University\\ $\{$koch,lantova$\}$@cs.cornell.edu
}
\date{}

\begin{document}

\maketitle

\vspace{-5mm}

\begin{abstract}
  Uncertain information is commonplace in real-world data management
  scenarios. The
  ability to represent large sets of possible instances (worlds) while
  supporting efficient storage and processing is an important challenge in
  this context.   The recent formalism of
  {\em world-set decompositions (WSDs)}\/
  provides a space-efficient representation
  for uncertain data that also supports scalable processing.  WSDs are {\em
    complete}\/ for finite world-sets in that they can represent any finite
  set of possible worlds.  For possibly infinite world-sets, we show that a
  natural generalization of WSDs precisely captures the expressive power of
  c-tables.  We then
  show that several important problems are
  efficiently solvable on WSDs while they are NP-hard on c-tables.  Finally,
  we give a polynomial-time algorithm for factorizing WSDs, i.e.\ an efficient
  algorithm for minimizing such representations.
\end{abstract}

\vspace{-7mm}

\section{Introduction}
\label{sec:introduction}

Recently there has been renewed interest in incomplete information databases.
This is due to the many important applications that systems for representing
incomplete information have, such as data cleaning, data integration, and
scientific databases.

{\em Strong representation
  systems}\/ \cite{IL1984,AKG1991,GT2006} are formalisms for
representing sets of possible worlds which are closed under query
operations in a given query language.  While there have been numerous other
approaches to dealing with incomplete information, such as closing
possible worlds semantics using certain answers
\cite{AD98,ABC2003,CDLR2004}, constraint or database repair
\cite{CMS2004,BFFR2005,BBFL2005}, and probabilistic ranked retrieval
\cite{dalvi04efficient,miller06clean}, strong representation systems
form a compositional framework that is minimally intrusive by not
requiring to lose information, even about the lack of information,
present in an information system: Computing certain answers, for example,
entails a loss of possible but uncertain information. Strong
representation systems can be nicely combined with the other
approaches. For example, data transformation queries and data cleaning
steps effected within a strong representation systems framework can be
followed by a query with ranked retrieval or certain answers
semantics, closing the possible worlds semantics.

The so-called {\em c-tables}\/ \cite{IL1984,Gra1984,Gra1991} are the
prototypical strong representation system. However, c-tables are not
well suited for representing large incomplete databases in
practice. Two recent works presented strong, indeed {\em complete},
representation systems for finite sets of possible worlds. The
approach of the {\em Trio x-relations}\/~\cite{trio} relies on a form
of intensional information (``lineage'') only in combination with
which the formalism is strong.
In~\cite{AKO06WSD} large sets of possible worlds are managed using
{\em world-set decompositions (WSDs)}\/. The approach is based on
relational product decomposition to permit space-efficient
representation. \cite{AKO06WSD} describes a prototype implementation
and shows the efficiency and scalability of the formalism in terms of
storage and query evaluation in a large census data scenario with up
to $2^{10^6}$ worlds, where each world stored is several GB in size.

\subsubsection*{Examples of world-set decompositions.}

\begin{figure}
  \begin{center}
    \begin{multicols}{2}
      \includegraphics[width=.45\textwidth]{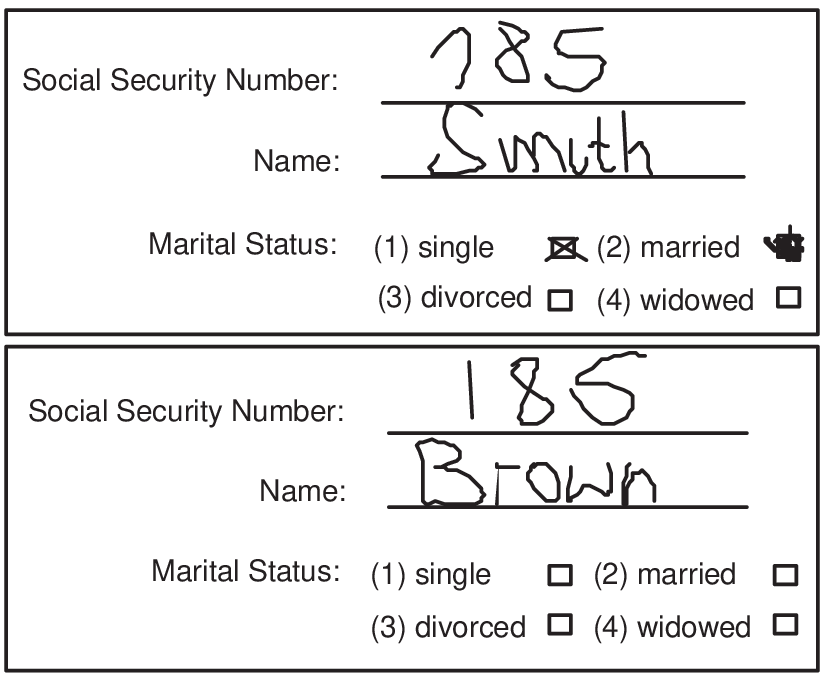}

    {\small
\begin{tabular}{cccccc}
$t_1.S$ & $t_1.N$ & $t_1.M$ &
$t_2.S$ & $t_2.N$ & $t_2.M$ \\
\hline
185 & Smith & 1 & 186 & Brown & 1 \\
185 & Smith & 1 & 186 & Brown & 2 \\
185 & Smith & 1 & 186 & Brown & 3 \\
185 & Smith & 1 & 186 & Brown & 4 \\
185 & Smith & 2 & 186 & Brown & 1 \\
185 & Smith & 2 & 186 & Brown & 2 \\
185 & Smith & 2 & 186 & Brown & 3 \\
185 & Smith & 2 & 186 & Brown & 4 \\
\multicolumn{6}{c}{\vdots}\\
\nop{ 
785 & Smith & 1 & 185 & Brown & 1 \\
785 & Smith & 1 & 185 & Brown & 2 \\
785 & Smith & 1 & 185 & Brown & 3 \\
785 & Smith & 1 & 185 & Brown & 4 \\
785 & Smith & 1 & 186 & Brown & 1 \\
785 & Smith & 1 & 186 & Brown & 2 \\
785 & Smith & 1 & 186 & Brown & 3 \\
785 & Smith & 1 & 186 & Brown & 4 \\
785 & Smith & 2 & 185 & Brown & 1 \\
785 & Smith & 2 & 185 & Brown & 2 \\
785 & Smith & 2 & 185 & Brown & 3 \\
785 & Smith & 2 & 185 & Brown & 4 \\
785 & Smith & 2 & 186 & Brown & 1 \\
785 & Smith & 2 & 186 & Brown & 2 \\
785 & Smith & 2 & 186 & Brown & 3 \\
}
785 & Smith & 2 & 186 & Brown & 4 \\
\hline
\end{tabular}
}
\end{multicols}
\end{center}

\vspace{-4mm}

\caption{Two completed survey forms and a world-set relation representing 
  the possible worlds with unique social security numbers.}
\label{fig:census}
\end{figure}

As WSDs play a central role in this work, we next exemplify them using two
manually completed forms that may originate from a census and which
allow for more than one interpretation (Figure~\ref{fig:census}). For
simplicity we assume that social security numbers consist of only
three digits. For instance, Smith's social security number can be read
either as ``185'' or as ``785''. We can represent the available
information using a relation in which possible alternative values are
represented in set notation (so-called or-sets): \vspace{-0.1cm}
{\footnotesize
\begin{center}
\begin{tabular}{c|ccc}
(TID) & S & N & M \\
\hline
$t_1$ & ~\{ 185, 785 \}~ & Smith & ~\{ 1, 2 \} \\
$t_2$ & ~\{ 185, 186 \}~ & Brown & ~\{ 1, 2, 3, 4 \} \\
\end{tabular}
\end{center}
}
\vspace{-0.1cm}

This or-set relation represents
$2\cdot2\cdot2\cdot4 = 32$ possible worlds.

We now enforce the integrity constraint that all social security
numbers be unique. For our example database, this constraint excludes
8 of the 32 worlds, namely those in which both tuples have the value
185 as social security number. This constraint excludes the worlds in
which both tuples have the value 185 as social security number.  It is
impossible to represent the remaining 24 worlds using or-set
relations. What we could do is store each world explicitly using a
table called a {\em world-set relation}\/ of a given set of worlds.
Each tuple in this table represents one world and is the concatenation
of all tuples in that world (see Figure \ref{fig:census}).

A world-set decomposition is a decomposition of a world-set relation
into several relations such that their product (using the product
operation of relational algebra) is again the world-set relation.  The
world-set represented by our initial or-set relation can also be
represented by the product

{\footnotesize
\[
\hspace{-1em}
\begin{tabular}{|c|}
\hline
$t_1.S$ \\
\hline
185 \\
785 \\
\hline
\end{tabular}
\times
\begin{tabular}{|c|}
\hline
$t_1.N$ \\
\hline
Smith \\
\hline
\end{tabular}
\times
\begin{tabular}{|c|}
\hline
$t_1.M$ \\
\hline
1 \\
2 \\
\hline
\end{tabular}
\times
\begin{tabular}{|c|}
\hline
$t_2.S$ \\
\hline
185 \\
186 \\
\hline
\end{tabular}
\times
\begin{tabular}{|c|}
\hline
$t_2.N$ \\
\hline
Brown \\
\hline
\end{tabular}
\times
\begin{tabular}{|c|}
\hline
$t_2.M$ \\
\hline
1 \\
2 \\
3 \\
4 \\
\hline
\end{tabular}
\]
}

In the same way we can represent the result of data cleaning with the
uniqueness constraint for the social security numbers as the product

{\footnotesize
\[
\begin{tabular}{|c@{\extracolsep{1mm}}c|}
\hline
$t_1.S$ & $t_2.S$ \\
\hline
185 & 186 \\
785 & 185 \\
785 & 186 \\
\hline
\end{tabular}
\times
\begin{tabular}{|c|}
\hline
$t_1.N$ \\
\hline
Smith \\
\hline
\end{tabular}
\times
\begin{tabular}{|c|}
\hline
$t_1.M$ \\
\hline
1 \\
2 \\
\hline
\end{tabular}
\times
\begin{tabular}{|c|}
\hline
$t_2.N$ \\
\hline
Brown \\
\hline
\end{tabular}
\times
\begin{tabular}{|c|}
\hline
$t_2.M$ \\
\hline
1 \\
2 \\
3 \\
4 \\
\hline
\end{tabular}
\]
}

One can observe that the result of this product is exactly the world-set
relation in Figure \ref{fig:census}.
The decomposition is based on the {\em independence}\/ between sets of
fields, subsequently called {\em components}\/.  Only fields that
depend on each other, for example $t_1.S$ and $t_2.S$, belong to the
same component. Since $\{ t_1.S, t_2.S \}$ and $\{ t_1.M \}$ are
independent, they are put into different components.

WSDs can be naturally viewed as c-tables whose formulas have been put
into a {\em normal form}\/ represented by the component relations.
The following c-table with global condition $\phi$ is equivalent to
the WSD with our integrity constraint enforced.

{\footnotesize
  \begin{center}
      \begin{tabular}{c|c@{\extracolsep{1mm}}c@{\extracolsep{1mm}}cl}
        T & S & N & M & \textit{cond}\\
        \hline
          &   &   &   & 
            $\phi = ((x = 185 \land z = 186) \lor (x = 785 \land z = 185) \lor$ \\
          &   &   &   & 
            \hspace*{2em} $(x = 785 \land z = 186)) \land (y = 1 \lor y = 2)\land$\\
          &   &   &   & 
            \hspace*{2em} $(w = 1 \lor w = 2 \lor w = 3 \lor w = 4)$\\\hline
        & $x$ & Smith & $y$ & \\
        & $z$ & Brown & $w$ & \\
      \end{tabular}

\end{center}
}

Formal definitions of WSDs and c-tables will be given in the body of this
article.

\subsubsection*{Contributions.}

\begin{table}[t]
  \begin{center}
    \framebox[.95\textwidth]{
    \begin{tabular}{l@{\hspace*{1em}}ll}
      Input     & \multicolumn{2}{l}{Representation system ${\cal W}$, instance $I=(R^I)$, tuple $t$}\vspace*{1em}\\
      Problems  & Tuple possibility:    & $\exists {\cal A}\in\mathit{rep}({\cal W}): t\in R^{\cal A}$\\
                & Tuple certainty:      & $\forall {\cal A}\in\mathit{rep}({\cal W}): t \in R^{\cal A}$ \hspace*{1em}\\
                & Instance possibility: & $\exists {\cal A}\in\mathit{rep}({\cal W}): R^I = R^{\cal A}$\\
                & Instance certainty:   & $\forall {\cal A}\in\mathit{rep}({\cal W}): R^I = R^{\cal A}$ \hspace*{1em}\\
                & Tuple q-possibility (query q fixed):    & $\exists {\cal A}\in\mathit{rep}({\cal W}): t\in q({\cal A})$\\
                & Tuple q-certainty (query q fixed):      & $\forall {\cal A}\in\mathit{rep}({\cal W}): t\in q({\cal A})$ \hspace*{1em}\\
                & Instance q-possibility (query q fixed): & $\exists {\cal A}\in\mathit{rep}({\cal W}): R^I = q({\cal A})$ \hspace*{1em}\\
                & Instance q-certainty (query q fixed):   & $\forall {\cal A}\in\mathit{rep}({\cal W}): R^I = q({\cal A})$ \hspace*{1em}\\
    \end{tabular}}
\end{center}
\caption{Decision Problems for Representation Systems.}
\label{tab:problems}

\vspace{-6mm}

\end{table}

The main goal of this work is to develop expressive yet efficient
representation systems for infinite world-sets and to study the
theoretical properties (such as expressive power, complexity of
query-processing, and minimization) of these representation systems.
Many of these results also apply to -- and are new for -- the
world-set decompositions of \cite{AKO06WSD}.

In \cite{GT2006}, a strong argument is made supporting c-tables as a
benchmark for the expressiveness of representation systems; we concur.
Concerning efficient processing, we adopt a less expressive syntactic
restriction of c-tables, called v-tables~\cite{IL1984,AKG1991}, as a
lower bound regarding succinctness and complexity. The main
development of this article is a representation system that combines,
in a sense, the best of all worlds:
(1) It is just as expressive as c-tables,
%
(2) it is exponentially more succinct than unions of v-tables, and
%
(3) on most classical decision problems, the complexity bounds are not
worse than those for v-tables.

\medskip

In more detail, the technical contributions of this article are as
follows\footnote{ This article extends
\cite{AKO06iWSD} with proofs, a modified algorithm for relational factorization with 
better space complexity, and new data complexity results for tuple
q-possibility, tuple q-certainty, and instance q-certainty, where $q$
is a full or positive relational algebra query.}:
\begin{itemize}
\item We introduce gWSDs, an extension of the WSD model of \cite{AKO06WSD}
  with variables and possibly negated equality conditions.
    
\item We show that gWSDs are expressively equivalent to c-tables and
  are therefore a strong representation system for full relational
  algebra.
  
\item We study the complexity of the main data management
  problems~\cite{AKG1991,IL1984} regarding WSDs and gWSDs, summarized
  in Table~\ref{tab:problems}. Table~\ref{tab:decision-overview}
  compares the complexities of these problems in our context to those
  of existing strong representation systems like the well-behaved
  ULDBs of Trio\footnote{The complexity results for Trio are from
  \cite{trio} and were not verified by the authors.} and c-tables.

\item We present an efficient algorithm for optimizing gWSDs, i.e., for
  computing an equivalent gWSD whose size is smaller than that of a given
  gWSD. In the case of WSDs, this is a minimization algorithm that
  produces the unique maximal decomposition of a given WSD.
\end{itemize}

\begin{table}
  \centering
  \begin{tabular}{|l@{\hspace*{0.8em}}|@{\hspace*{0.8em}}l@{\hspace*{0.8em}}|@{\hspace*{0.8em}}l@{\hspace*{0.8em}}|@{\hspace*{0.8em}}l@{\hspace*{0.8em}}|@{\hspace*{0.8em}}l@{\hspace*{0.8em}}|}
    \hline
                           & v-tables~\cite{AKG1991}     & (g)WSDs   & Trio~\cite{trio} & c-tables~\cite{Gra1991}\\\hline\hline
    Tuple possibility      & PTIME       & PTIME         & PTIME            & NP-compl. \\\hline
    Tuple certainty        & PTIME       & PTIME         & PTIME            & coNP-compl. \\\hline
    Instance possibility   & NP-compl.   & NP-compl.     & NP-hard          & NP-compl.   \\\hline
    Instance certainty     & PTIME       & PTIME         & NP-hard          & coNP-compl.   \\\hline
    Tuple q-possibility    & NP-compl.   & NP-compl.     & ?                & NP-compl. \\
    \hspace*{.2em}\textit{positive relational algebra} & PTIME     & PTIME           & ?                & NP-compl. \\\hline
    Tuple q-certainty      & coNP-compl.  & coNP-compl.                 & ?                & coNP-compl. \\
    \hspace*{.2em}\textit{positive relational algebra} & PTIME       & coNP-compl.   & ?                & coNP-compl. \\\hline
    Instance q-possibility & NP-compl.   & NP-compl.     & NP-hard          & NP-compl. \\\hline
    Instance q-certainty   & coNP-compl. & coNP-compl.   & NP-hard          & coNP-compl. \\
    \hspace*{.2em}\textit{positive relational algebra}   & PTIME       & coNP-compl. & NP-hard          & coNP-compl. \\\hline
  \end{tabular}

\smallskip
  
\caption{Comparison of data complexities for standard decision problems.}
\label{tab:decision-overview}

\vspace{-8mm}

\end{table}

One can argue that gWSDs are a practically more applicable
representation formalism than c-tables: While having the same
expressive power, many important problems are easier to solve. Indeed,
as shown in Table~\ref{tab:decision-overview}, the complexity results
for gWSDs on many important decision problems are identical to those
for the much weaker v-tables.  At the same time WSDs are still concise
enough to support the space-efficient representation of very large
sets of possible worlds (cf.\ the experimental evaluation on WSDs in
\cite{AKO06WSD}). Also, while gWSDs are strictly stronger than Trio
representations, which can only represent finite world-sets, the
complexity characteristics are better.

The results on finding maximal product decompositions relate to earlier
work done by the database theory community on
relational decomposition given schema constraints (cf.\ e.g.\ \cite{AHV95}).
Our algorithms
do not assume such constraints and only take a snapshot of a database at a
particular point in time into consideration. Consequently, updates may require
to alter a decomposition. Nevertheless, our results may be of
interest  independently from WSDs
as for instance in certain scenarios with very dense relations, decompositions
may be a practically relevant technique for efficiently storing and querying
large databases.

Note that we do not consider probabilistic approaches to representing
uncertain data (e.g.\ the recent work \cite{dalvi04efficient}) in this
article.  However, there is a natural and straightforward
probabilistic extension of WSDs which directly inherits many of the
properties studied in this article, see \cite{AKO06WSD}.

The structure of the article basically follows the list of
contributions.


\vspace{-3mm}

\section{Preliminaries}
\label{sec:prel}

We use the named perspective of the relational model and relational algebra
with the operations selection $\sigma$, projection $\pi$, product $\times$,
union $\cup$, difference $- $, and renaming
$\delta$.

A {\em relation schema}\/ is a construct of the
form $R[U]$, where $R$ is a relation name and $U$ is a nonempty set of
attribute names.\footnote{For technical reasons involving the WSDs presented
later, we exclude nullary relations and will represent these (e.g., when
obtained as results from a Boolean query) using unary relations
over a special constant ``true''.}
Let ${\bf D}$ be an infinite set of atomic values,
the {\em domain}\/.  A {\em relation} over schema $R[A_1, \dots, A_k]$ is a
finite set of tuples $(A_1: a_1, \dots, A_k: a_k)$ where $a_1, \dots, a_k \in
{\bf D}$.  A {\em relational schema}\/ is a tuple $\Sigma = (R_1[U_1], \dots,
R_k[U_k])$ of relation schemas.
A {\em relational structure (or database)}\/ ${\cal A}$ over schema
$\Sigma$ is a tuple $(R_1^{\cal A}, \dots, R_k^{\cal A})$, where each
$R_i^{\cal A}$ is a relation over schema $R_i[U_i]$.
When no confusion may occur,
we will also use $R$ rather than $R^{\cal A}$ to denote one particular
relation over schema $R[U]$.  For a relation $R$, $\textsf{sch}(R)$ denotes
the set of its attributes, $ar(R)$ its
arity and $|R|$ the number of tuples in $R$.

A set of {\em possible worlds}\/ (or {\em world-set}) over schema
$\Sigma$ is a set of databases over schema $\Sigma$.
Let ${\bf W}$ be a set of finite structures, and let $rep$ be a function that
maps each ${\cal W} \in {\bf W}$ to a world-set of
the same schema. Then $({\bf W}, rep)$ is called a {\em strong representation
  system}\/
for a query language if, for each query $Q$ of that language and each
${\cal W} \in {\bf W}$ such that the schema of $Q$
is consistent with the schema of the worlds in
$rep({\cal W})$, there is a structure ${\cal W}' \in {\bf W}$ such
that $rep({\cal W}') = \{ Q({\cal A}) \mid {\cal A} \in rep({\cal W}) \}$.

\subsection{Tables}

We now review a number of representation systems for incomplete information
that are known from earlier work (cf.\ e.g.\ \cite{Gra1991,AHV95}).


Let ${\bf X}$ be a set of variables. We call an equality of the form
$x=c$ or $x=y$, where $x$ and $y$ are variables from ${\bf X}$ and $c$
is from ${\bf D}$ an {\em atomic condition}\/, and will define {\em
(general) conditions}\/ as Boolean combinations (using conjunction,
disjunction, and negation) of atomic conditions and the constant
``true''.

\begin{definition}[c-table]
\em
A {\em c-multitable}\/ \cite{IL1984,Gra1991} over schema
$(R_1[U_1], \dots, R_k[U_k])$ is a tuple
\[
{\cal T} = (R_1^{\cal T}, \dots, R_k^{\cal T}, \phi^{\cal T}, \lambda^{\cal T})
\]
where each $R_i^{\cal T}$ is a set of $ar(R_i)$-tuples over
${\bf D} \cup {\bf X}$, $\phi^{\cal T}$ is a  Boolean
combination over equalities on ${\bf D} \cup {\bf X}$ called the {\em global
  condition}, and function $\lambda^{\cal T}$ assigns
each tuple from one of the relations $R_1^{\cal T}, \dots, R_k^{\cal T}$ to
a condition (called the {\em local condition}\/ of the tuple).
A c-multitable with $k=1$ is called a {\em c-table}\/.

The semantics of a c-multitable ${\cal T}$, called its {\em representation}\/
$rep({\cal T})$, is defined via the notion of a valuation of the variables
occurring in ${\cal T}$ (i.e., those in the tuples as well as those in
the conditions). Let
$
\nu: {\bf X} \rightarrow {\bf D}
$
be a valuation that assigns each variable in ${\cal T}$ to a domain value.
We overload $\nu$ in the natural way to map tuples and conditions over
${\bf D} \cup {\bf X}$ to tuples and formulas over ${\bf D}$.\footnote{Done by
extending $\nu$ to be the identity on domain values and to commute with the
tuple constructor, the Boolean operations, and equality.}
A {\em satisfaction}\/ of ${\cal T}$ is a valuation $\nu$ such that
$\nu(\phi^{\cal T})$ is true. A satisfaction $\nu$
takes ${\cal T}$ to a relational structure
$\nu({\cal T}) = (R_1^{\nu({\cal T})}, \dots, R_k^{\nu({\cal T})})$
where each relation $R_i^{\nu({\cal T})}$ is obtained as
$
R_i^{\nu({\cal T})} := \{ \nu(\tv) \mid \tv \in R_i^{\cal T} \land
   \nu(\lambda^{\cal T}(\tv)) \mbox{ is true} \}.
$
The representation of ${\cal T}$ is now given by its satisfactions,
$
rep({\cal T}) := \{ \nu({\cal T}) \mid \nu \mbox{ is a satisfaction of
${\cal T}$} \}.$
%
%
\punto
\end{definition}

\begin{example}
Section~\ref{sec:introduction} gives a c-table $T$ representing our
uncertain census data of Figure~\ref{fig:census}. $T$ uses one
variable per uncertain field and lists the possible values of the
variables in the global condition $\phi$. Each satisfaction of $T$
defines a world and there are 24 such worlds. The local conditions in
$T$ are ``true'' and omitted.

Figure~\ref{fig:ex-ctab-wsd}(a) shows a c-table $T$, where both tuples
have local conditions. $T$ has infinitely many satisfactions and thus
defines an infinite world-set. For example, the satisfaction
$\{x\mapsto 2,y\mapsto 1,z\mapsto 2\}$ defines the world ${\cal A}$
with relation $T^{\cal A} = \{\nu(\tuple{A:x, B:1}) \mid
\nu(x\not=2) \mbox{ is true })\}\cup \{\nu(\tuple{A:z, B:y}) \mid
\nu(y\not=2) \mbox{ is true })\} = \{\tuple{A:2, B:1}\}$.\punto
\end{example}

\begin{proposition}[\cite{IL1984}]\label{prop:ctables-strong}
The c-multitables are a strong representation system for relational algebra.
\end{proposition}

We consider two important restrictions of c-multitables.
\begin{enumerate}
\item
By a {\em g-multitable}\/ \cite{AKG1991}, we refer to a c-multitable
in which the global condition $\phi^{\cal T}$is a conjunction of
possibly negated equalities and $\lambda^{\cal T}$ maps each tuple to
``true''.

\item
A {\em v-multitable} is a g-multitable in which the global condition
$\phi^{\cal T}$ is a conjunction of equalities.
\end{enumerate}

Without loss of generality, we may assume that the global condition of
a g-multitable is a conjunction of {\em negated equalities} and the
global condition of a v-multitable is simply ``true''.\footnote{Each
g-multitable resp.\ v-multitable can be reduced to one in this normal
form by variable replacement and the removal of tautologies such as
$x=x$ or $1=1$ from the global condition.}  Subsequently, we will
always assume these two normal forms and omit local conditions from
g-multitables and both global and local conditions from v-multitables.

\begin{figure}
\[
\begin{array}{cp{10mm}cp{10mm}c}
\phi^{\cal T} = (x \neq y)
\\[1mm]
\begin{array}{l|ll}
R^{\cal T} & A & B \\
\hline
  & x & 1 \\
  & 2 & x \\
\end{array}
\hspace{5mm}
\begin{array}{l|l}
S^{\cal T} & C \\
\hline
  & y \\
  & 3 \\
\end{array}
&&
\begin{array}{l|ll}
R{\cal A} & A & B \\
\hline
  & 1 & 1 \\
  & 2 & 1 \\
\end{array}
\hspace{5mm}
\begin{array}{l|l}
S{\cal A} & C \\
\hline
  & 2 \\
  & 3 \\
\end{array}
&&
\nu: \left\{
\begin{array}{lll}
x & \mapsto & 1 \\
y & \mapsto & 2 \\
\end{array}
\right.
\\[6mm]
(a) && (b) && (c)
\end{array}
\]
\caption{A g-multitable ${\cal T}$ (a), possible world ${\cal A}$ (b),
and a valuation s.t.\  $\nu({\cal T}) = {\cal A}$ (c).}
\label{fig:itable}
\vspace*{-2em}
\end{figure}

\begin{example}
Consider the g-multitable ${\cal T} = (R^{\cal T}, S^{\cal T},
\phi^{\cal T})$ of Figure~\ref{fig:itable}~(a). Then the valuation of
Figure~\ref{fig:itable}~(c) satisfies the global condition of ${\cal
T}$, as $\nu(x) \neq \nu(y)$.  Thus ${\cal A} \in rep({\cal T})$,
where ${\cal A}$ is the structure from
Figure~\ref{fig:itable}~(b).\punto
\end{example}

\begin{remark}
It is known from \cite{IL1984} that v-tables are not a strong representation
system for relational selection, but for the fragment of relational algebra
built from projection, product, and union.

The definition of c-multitables used here is from \cite{Gra1991}. The original
definition from \cite{IL1984} has been more restrictive in
requiring the global condition to be ``true''.
While c-tables without a global condition are strictly weaker
(they cannot represent the empty world-set),
they nevertheless form a strong representation system for relational
algebra.

In \cite{AHV95}, the global conditions of c-multitables are required to be
conjunctions of possibly negated equalities. It will be a corollary of a
result of this paper (Theorem~\ref{th:equiv}) that this definition is
equivalent to c-multitables with arbitrary global conditions.  \punto
\end{remark}

We next define a restricted form of c-tables, called mutex-tables (or
x-tables for short). This formalism is of particular importance in
this paper as it is closely related to gWSDs, our main representation
formalism. An x-table is a c-table where the global condition is a
conjunction of negated equalities and the local conditions are
conjunctions of equalities and a special form of negated
equalities. We make this more precise next.

Consider a set of variables ${\bf Y}$ and a function $\mu:{\bf
  Y}\mapsto\mathbb{N}^+$ mapping variables to positive numbers. The
\textit{mutex set} $\mathbb{M}({\bf Y},\mu)$ for ${\bf Y}$ and $\mu$
is defined by
$$\{\mathrm{``true''}\}\cup\{(x=i)\mid x\in{\bf Y}, 1\leq
i\leq\mu(x)\}\cup\{(x\not=1\wedge\ldots\wedge x\not=\mu(x))\mid x\in{\bf
  Y}\}.$$
Intuitively, $\mathbb{M}$ defines for each variable of ${\bf Y}$
possibly negated equalities such that a variable valuation satisfies precisely
one of these conditions.

\begin{definition}[x-table]\label{def:xtables}
  \em An {\em x-multitable} is a c-multitable $${\cal T} = (R_1^{\cal
  T}, \dots, R_k^{\cal T}, \phi^{\cal T}, \lambda^{\cal T}),$$ where
  (1) the global condition $\phi^{\cal T}$ is a conjunction of negated
  equalities, (2) all local conditions defined by $\lambda^{\cal T}$
  are conjunctions over formulas from a mutex set $\mathbb{M}({\bf
  Y},\mu)$ and equalities over ${\bf X}\cup{\bf D}$, and (3) the
  variables in ${\bf Y}$ do not occur in $R_1^{\cal T}, \dots,
  R_k^{\cal T}, \phi^{\cal T}$. An x-multitable with $k=1$ is called
  an {\em x-table}\/.  \punto
\end{definition}

\begin{example}
Figure~\ref{fig:ex-wsd-ctab-b} shows an x-table $T$ over the mutex set
$\mathbb{M}({\bf Y},\mu)$ where ${\bf Y}=\{x_1\}$ and $\mu(x)=1$. The
mutex conditions on $x_1$ are used to state that instantiations of the
first tuple cannot occur in the same worlds with instantiations of the
last two tuples.

Figure~\ref{fig:decision-tuple}(b) shows an x-multitable ${\cal T}$
over a mutex set with ${\bf Y}=\{x_1,x_3\}$ and
$\mu(x_1)=\mu(x_3)=1$. The mutex conditions on $x_1$ are used to state
that instantiations of the first two tuples of $R$ and of the first
tuple of $S$ cannot occur in the same worlds with instantiations of
the third tuple of $R$ and the second tuple of $S$. For example, the
satisfaction $\{x_1\mapsto 2,x_3\mapsto 2,y\mapsto 3,z\mapsto 4\}$ of
${\cal T}$ defines the world ${\cal A}$ with $R^{\cal
A}=\{\tuple{A:2},\tuple{A:1}\}$ and $S^{\cal A}=\{\tuple{B:2}\}$,
whereas the satisfaction $\{x_1\mapsto 1,x_3\mapsto 1,y\mapsto
3,z\mapsto 4\}$ defines the world ${\cal B}$ with $R^{\cal
B}=\{\tuple{A:2},\tuple{A:3},\tuple{A:1}\}$ and $S^{\cal
B}=\{\tuple{B:4},\tuple{B:1}\}$.\punto
\end{example}

It will be a corollary of joint results of this paper
(Lemma~\ref{lem:wsdx-c} and Theorem~\ref{th:equiv}) that x-multitables
are as expressive as c-multitables.

\begin{proposition}
  The x-multitables capture the c-multitables.
\end{proposition}

This result implies that x-multitables are a strong representation
system for relational algebra. In this paper, however, we will make
particular use of a weaker form of strongness, namely for positive
relational algebra, in conjunction with efficient query evaluation.

\begin{proposition}\label{prop:xtables}
  The x-multitables are a strong representation system for positive
  relational algebra. The evaluation of positive relational algebra
  queries on x-multitables has polynomial data complexity.
  \begin{proof}\em

    We use the algorithm of \cite{IL1984,Gra1991} for the evaluation
    of relational algebra queries on c-multitables and obtain an
    answer c-multitable of polynomial size. Consider a fixed positive
    relational algebra query $Q$, c-multitable ${\cal T}$, and c-table
    ${\cal T}'$, where ${\cal T}'$ represents the answer to $Q$ on
    ${\cal T}$. We compute ${\cal T}'$ by recursively applying each
    operator in $Q$. The evaluation follows the relational case except
    for the computation of global and local conditions (which do not
    exist in the relational case). The global condition of ${\cal T}$
    becomes the global condition of ${\cal T}'$. For projection and
    union, tuples preserve their local conditions from the input. In
    case of selection, the local condition of a result tuple is the
    conjunction of the local condition of the input tuple and, if
    required by the selection condition, of new equalities involving
    variables in the tuple and constants from the positive selections
    of $Q$. In case of product, the local condition of a result tuple
    is the conjunction of the local conditions of the constituent
    input tuples.
    
    The local conditions in ${\cal T}'$ are thus conjunctions of local
    conditions of ${\cal T}$ and possibly additional equalities. In case
    ${\cal T}$ is an x-table, then its local conditions are conjunctions over
    formulas from a mutex set $\mathbb{M}$ and further equalities. Thus the
    local conditions of ${\cal T}'$ are also conjunctions over formulas from
    $\mathbb{M}$ and further equalities. ${\cal T}'$ is then an x-table.
    \punto
  \end{proof}
\end{proposition}





\vspace{-3mm}

\section{New Representation Systems}
\label{sec:wsd}

This section introduces novel representation systems beyond those
surveyed in the previous section. We start with finite sets of
v(g-,c-)tables, or tabsets for short, then show how to inline tabsets
into tabset-tables, and finally introduce decompositions of such
tabset-tables based on relational product. Such decompositions are our
main vehicle for representing incomplete data and the next sections
are dedicated to their expressiveness and efficiency.

\subsection{Tabsets and Tabset Tables}

We consider finite sets of multitables as representation systems, and
will refer to such constructs as {\em tabsets}\/ (rather than as {\em
multitable-sets}\/, to be short).

A g-(resp., v-)tabset ${\bf T} = \{{\cal T}_1,\ldots, {\cal T}_n
\}$ is a finite set of g-(v-)multitables.
The representation of a tabset is the union of the representations of
the constituent multitables,
\[
rep({\bf T}) :=
rep({\cal T}_1) \cup \dots \cup rep({\cal T}_n).
\]

Note that finite sets of v-multitables are more expressive than
v-multitables: v-tabsets can trivially represent any finite world-set
with one v-multitable representing precisely one world. It is
known~\cite{AHV95} that no v-multitable can represent the world-set
consisting of an empty world and a non-empty world, as produced by,
e.g., selection queries on v-multitables.






We next construct an {\em inlined}\/ representation of a tabset as a
single table by turning each multitable into a single tuple.

Let {\bf A} be a g-tabset over schema $\Sigma$. For each $R[U]$ in
$\Sigma$, let $|R|_{\max} = \max \{ |R^{\cal A}| : {\cal A} \in {\bf A}
\}$ denote the maximum cardinality of $R$ in any multitable of ${\bf
A}$. Given a g-multitable ${\cal A} \in {\bf A}$ with $R^{\cal A} =
\{ \tv_1, \dots, \tv_{|R^{\cal A}|} \}$, let $\mbox{inline}(R^{\cal
A})$ be the tuple obtained as the concatenation (denoted $\circ$) of
the tuples of $R^{\cal A}$ padded with a special tuple $t_\bot$ up to
arity $|R|_{\max}$,
\[
\mbox{inline}(R^{\cal A}) :=
\tv_1 \circ \dots \circ \tv_{|R^{\cal A}|} \circ
(\underbrace{t_{\bot}, \dots\dots\dots, t_{\bot}}_{|R|_{\max} -
  |R^{\cal A}|}), \mbox{ where } t_{\bot} = \tuple{\underbrace{\bot,\ldots,\bot}_{ar(R)}}
\]
Then tuple
\[
\mbox{inline}({\cal A}) :=
\mbox{inline}(R_1^{\cal A}) \circ \dots \circ \mbox{inline}(R_{|\Sigma|}^{\cal A})
\]
encodes all the information in ${\cal A}$.

We make use of the symbol $\bot$ to align the g-tables of different
sizes and uniformly inline g-tabsets. Given a g-multitable ${\cal A}$
padded with additional tuples $t_{\bot}$, there is no world
represented by $\mbox{inline}({\cal A})$ that contains instantiations
of these tuples. We extend this interpretation and generally define as
$t_{\bot}$ any tuple that has at least one symbol $\bot$, i.e.,
$\tuple{A_1:a_1,\ldots,A_n:a_n}$, where at least one $a_i$ is $\bot$,
is a $t_{\bot}$ tuple. This allows for several different inlinings
that represent the same world-set.

\begin{definition}[gTST]
\em
Given an inlining function \mbox{inline}, a {\em g-tabset table
(gTST)}\/ of a g-tabset ${\bf A}$ is the pair $(W, \lambda)$
consisting of the table\footnote{Note that this table may contain
variables and occurrences of the $\bot$ symbol.} $ W =
\{\mbox{inline}({\cal A}) \mid {\cal A} \in {\bf A} \} $ and the
function $\lambda$ which maps each tuple $\mbox{inline}({\cal A})$ of
$W$ to the global condition of ${\cal A}$.
\punto
\end{definition}

A vTST (TST) is obtained in strict analogy, omitting $\lambda$
($\lambda$ and variables).

To compute $\mbox{inline}(R^{\cal A})$, we have fixed an arbitrary
order of the tuples in $R^{\cal A}$. We represent this order by using
indices $\tid_i$ to denote the $i$-th tuple in $R^{\cal A}$ for each
g-multitable ${\cal A}$, if that tuple exists.  Then the TST has
schema
\[
\{ R.\tid_i.A_j \mid R[U] \textrm{ in } \Sigma, 1 \le i \le |R|_{\max}, A_j \in U \}.
\]

\begin{example}
An example translation from a tabset to a TST is given in
Figure~\ref{fig:itst}.
\end{example}

\begin{figure}[t]
\begin{center}
\[
\begin{array}{cp{8mm}cp{8mm}c}
\phi^{{\cal A}} && \phi^{{\cal B}} && \phi^{{\cal C}} \\[1mm]
\begin{array}{l|ll}
R^{\cal A} & A & B \\
\hline
& a_1 & a_2 \\
& a_3 & a_4 \\
\\
\end{array}
\hspace{5mm}
\begin{array}{l|ll}
S^{\cal A} & C \\
\hline
& a_5 \\
& a_6 \\
\\
\end{array}
&&
\begin{array}{l|ll}
R^{\cal B} & A & B \\
\hline
& b_1 & b_2 \\
& b_3 & b_4 \\
& b_5 & b_6
\end{array}
\hspace{5mm}
\begin{array}{l|ll}
S^{\cal B} & C \\
\hline
\\
\\
\\
\end{array}
&&
\begin{array}{l|ll}
R^{\cal C} & A & B \\
\hline
& c_1 & c_2 \\
\\
\\
\end{array}
\hspace{5mm}
\begin{array}{l|ll}
S^{\cal C} & C \\
\hline
& c_3 \\
& c_4 \\
& c_5
\end{array}
\end{array}
\]
(a) Three $(R[A,B], S[C])$-multitables ${\cal A}$, ${\cal B}$, and ${\cal C}$.

\[
\begin{array}{l|cccccc|ccc|l}
& R.\tid_1.A & R.\tid_1.B & R.\tid_2.A & R.\tid_2.B & R.\tid_3.A & R.\tid_3.B
& S.\tid_1.C & S.\tid_2.C & S.\tid_3.C & \lambda \\
\hline
& a_1 & a_2 & a_3  & a_4  & \bot & \bot & a_5  & a_6  & \bot & \phi^{{\cal A}} \\
& b_1 & b_2 & b_3  & b_4  & b_5  & b_6  & \bot & \bot & \bot & \phi^{{\cal B}} \\
& c_1 & c_2 & \bot & \bot & \bot & \bot & c_3  & c_4  & c_5  & \phi^{{\cal C}} \\
\end{array}
\]
(b): TST of tabset $\{ {\cal A}, {\cal B}, {\cal C} \}$.
\end{center}

\vspace{-5mm}

\caption{Translation from a tabset (a) to a TST (b).}
\label{fig:itst}

\vspace{-5mm}

\end{figure}

The semantics of a gTST $(W, \lambda)$ as a representation system is given in
strict analogy with tabsets,
\[
rep(W, \lambda) := \bigcup \{ rep(\mbox{inline}^{-1}(\tv), \lambda(\tv))
   \mid \tv \in W \}.
\]

\begin{remark}
Computing the inverse of ``inline''
is an easy exercise.
In particular, we map
$\mbox{inline}(R^{\cal A})$
to $R^{\cal A}$ as
\begin{multline*}
(a_1, \dots, a_{ar(R) \cdot |R|_{\max}})
\mapsto
\{
(a_{ar(R) \cdot k + 1}, \dots, a_{ar(R) \cdot (k+1)}) \mid
   0 \le k < |R|_{\max}, \\
   a_{ar(R) \cdot k + 1} \neq \bot, \dots, a_{ar(R) \cdot (k+1)}  \neq \bot
\}.
\end{multline*}
\end{remark}

By construction, the TSTs capture the tabsets.

\begin{proposition}
The gTSTs (resp., vTSTs) capture the g-(v-)tabsets.
\end{proposition}

Finally, there is a noteworthy normal form for gTSTs.

\begin{proposition}
The gTST in which $\lambda$ maps each tuple to a common global
condition $\phi$ unique across the gTST, that is,
$\lambda: \cdot \mapsto \phi$, capture the gTST.
\end{proposition}

\begin{proof}
Given a g-tabset ${\bf A}$, we may assume without loss of generality
that no two
g-multitables from ${\bf A}$ share a common variable, either in the
tables or the conditions, and that all global conditions in ${\bf A}$
are satisfiable. (Otherwise we could safely remove some of the
g-multitables in ${\bf A}$.) But, then, $\phi$ is simply the
conjunction of the global conditions in ${\bf A}$.  For any tuple
$\tv$ of the gTST of ${\bf A}$, the g-multitable
$(\mbox{inline}^{-1}(\tv), \phi)$ is equivalent to
$(\mbox{inline}^{-1}(\tv),
\lambda(\tv))$.
\punto
\end{proof}

\begin{proviso}
\em
We will in the following write gTSTs as pairs $(W, \phi)$, where $W$
is the table and $\phi$ is a single global condition shared by the
tuples of $W$.
\end{proviso}


\subsection{World-set Decompositions}
\label{sec:wsds}

We are now ready to define world-set decompositions,
our main vehicle for efficient yet
expressive representation systems.

A {\em product $m$-decomposition} of a relation $R$ is a set of
non-nullary relations $\{C_1,\ldots,C_m\}$ such that
$C_1\times\cdots\times C_m = R$. The relations $C_1,\ldots,C_m$ are
called {\em components}. A product $m$-decomposition of $R$ is {\em
  maximal(ly decomposed)}\/ if there is no product $n$-decomposition of $R$
with $n>m$.

\begin{definition}[attribute-level gWSD]
\em
Let $(W, \phi)$ be a gTST.  Then an {\em attri\-bute-level world-set
$m$-decomposition ($m$-gWSD)}\/ of $(W, \phi)$ is a pair of a product
$m$-decomposition of $W$ together with the global condition $\phi$.
\punto
\end{definition}

We also consider two important simplifications of gWSDs, those without
global condition, called vWSDs, and vWSDs without variables, called
WSDs. An example of a WSD is shown in Figure~\ref{fig:wsdex}.

\begin{figure}[t]
\[
\left\{
\begin{array}{l|cc}
R & A & B \\
\hline
\tid_1 & 1 & 2 \\
\tid_2 & 5 & 6 \\
\end{array}
\hspace{3mm}
\begin{array}{l|cc}
R & A & B \\
\hline
\tid_1 & 1 & 2 \\
\\
\end{array}
\hspace{3mm}
\begin{array}{l|cc}
R & A & B \\
\hline
\tid_1 & 3 & 4 \\
\tid_2 & 5 & 6 \\
\end{array}
\hspace{3mm}
\begin{array}{l|cc}
R & A & B \\
\hline
\tid_1 & 3 & 4 \\
\\
\end{array}
\right\}
\hspace{10mm}
\begin{array}{l|cccc}
C_1 & R.\tid_1.A & R.\tid_1.B \\
\hline
& 1 & 2 \\
& 3 & 4 \\
\end{array}
\times
\begin{array}{l|cccc}
C_2 & R.\tid_2.A & R.\tid_2.B \\
\hline
& 5 & 6 \\
& \bot & \bot \\
\end{array}
\]

\vspace{-4mm}

\caption{Set of four worlds and a corresponding 2-WSD.}
\label{fig:wsdex}

\vspace{-4mm}

\end{figure}

The semantics of a gWSD is given by its exact correspondence with
a gTST,
\[
rep\underbrace{(\{ C_1, \dots, C_m \}, \phi)}_{\mathrm{gWSD}} :=
rep\underbrace{(C_1 \times \dots \times C_m, \phi)}_{\mathrm{gTST}}.
\]

To decompose $W$, we treat its variables and the $\bot$-value as
constants. Clearly, the g-tabset ${\bf A}$ and any gWSD of ${\bf A}$
represent the same set of possible worlds.


It immediately follows from the definition of WSDs that

\begin{proposition}
\label{prop:finite_strong}
Any finite set of possible worlds can be represented as a $1$-WSD.
\end{proposition}
\begin{corollary}\label{cor:finite_strong}
WSDs are a strong re\-pre\-sen\-tation system for any relational
query language.
\end{corollary}

In the case of infinite world-sets, however, the mere extension of
WSDs with variables and equalities does not suffice to make them
strong. The lack of power to express negated equalities, despite the
ability to express disjunction, keeps vWSDs (and thus equally
v-tabsets) from being strong in the case of infinite world-sets.

\begin{proposition}
\label{prop:wsdx}
vWSDs are a strong representation system for projection, product and
union, and are not a strong representation system for selection and
difference.

\begin{proof}
\em
We show that v-tabsets are a strong representation system for
projection, product and union but not for selection and difference.
From the equivalence of v-tabsets and vWSDs (each v-tabset is a
1-vWSD) the property also holds for vWSDs.

Let ${\cal T} = \{{\cal T}_1, \ldots, {\cal T}_n\}$ be a v-tabset of
multitables over schema $\Sigma$. The results of the operations
projection $\pi_U(R_1)$, product $R_1 \times R_2$ and union $R_1 \cup
R_2$ on ${\cal T}$, respectively, (with $R_1, R_2\in\Sigma$) are then
defined as
\begin{align*}
\pi_U(R_1)({\cal T})       &= \{R'\ |\ {\cal T}_i \in {\cal T}, R' =  \pi_U(R_1^{{\cal T}_i}) \}\\
(R_1 \cup R_2)({\cal T})   &= \{R'\ |\ {\cal T}_i \in {\cal T}, R' =  R_1^{{\cal T}_i} \cup R_2^{{\cal T}_i}\}\\
(R_1 \times R_2)({\cal T}) &= \{R'\ |\ {\cal T}_i \in {\cal T}, R' =  R_1^{{\cal T}_i} \times R_2^{{\cal T}_i}\}
\end{align*}

To show that v-tabsets are not strong for selection and difference we
consider a v-tabset consisting of the following v-multitable $(R,S)$:
\begin{center}
\begin{tabular}{c|cc}
                $R$ & $A$ & $B$ \\\hline
                $\tid_1$ & $x$ & 2\\
                $\tid_2$ & 1   & $x$\\
\end{tabular}
\hspace{1cm}
\begin{tabular}{c|cc}
                $S$ & $A$ & $B$ \\\hline
                $\tid_3$ & 1 & 1\\
\end{tabular}
\end{center}

Consider the selection $\sigma_{A=1}(R)$. The answer world-set $W$
consists of the world $\{\tuple{A : 1,B : 2},\tuple{A : 1,B : 1}\}$ in
case $x=1$, and the worlds $\{\tuple{A : 1,B : c}\}$, where $c\in{\bf
D}-\{1\}$, in case $x\not=1$. We prove by contradiction that there is
no v-tabset representing precisely the world-set $W$. Since $W$ is an
infinite world-set and a v-tabset consists of only finitely many
v-tables, there must be at least one v-table $T$ that represents
infinitely many worlds of the form $\{ \tuple{A:1, B:c} \mid c \in
D\}$ and $rep(T) \subseteq W$. Since all tuples in a world of $W$ have
1 as a value for $A$, all tuples in $T$ must have it too, otherwise
$T$ will represent worlds that are not in $W$. Also, to represent
infinitely many worlds, $T$ must contain at least one variable. Thus
$T$ consists of v-tables with tuples of the form $\tuple{A: 1, B: y}$,
where for at least one such tuple $y$ is a variable. But then for
$y\mapsto 3$, a v-table containing $\tuple{A: 1, B: y}$ with variable
$y$ must not contain any other tuple whose instantiation is different
from $\tuple{A: 1, B: 3}$, as there are no worlds in $W$ containing
$\tuple{A: 1, B: 3}$ and other different tuples. This implies that for
$y\mapsto 1$, $W$ has either a world $\{\tuple{A: 1, B: 1}\}$ (in case
of v-tables with one tuple $\tuple{A: 1, B: y}$), or a world
$\{\tuple{A: 1, B: 1},\tuple{A: 1, B: 3}\}$ (in case of v-tables with
several more tuples). Contradiction.

Consider now the difference $R - S$. The answer world-set $W'$
consists of the world $\tuple{A:1, B:2}$ in case $x=1$, and the worlds
$\{\tuple{A:c,B:2}, \tuple{A:1, B:c}\}$, $c\in{\bf D}-\{1\}$, in case
$x\not=1$. We prove by contradiction that there is no v-tabset
representing precisely the world-set $W$. Using arguments similar to
the above case of selection, the answer v-tabset consists of v-tables
that have (possibly many) tuples of the form $\{\tuple{A:y,B:2},
\tuple{A:1, B:y}\}$, where $y$ is a variable for at least one pair of
such tuples. But then, for $y\mapsto 1$, there are worlds that contain
$\{\tuple{A:1,B:2},
\tuple{A:1, B:1}\}$ and these worlds are not in $W$. Contradiction.\punto
\end{proof}
\end{proposition}
We will later see that, in contrast to vWSDs, gWSDs are a strong
representation system for any relational language, because they
capture c-multitables (Theorem~\ref{th:equiv}).

\begin{remark}
Verifying nondeterministically that a structure ${\cal A}$ is a
possible world of gWSD $(\{ C_1, \dots, C_m \}, \phi)$ is easy: all we
need is choose one tuple from each of the component tables $C_1,
\dots, C_m$, concatenate them into a tuple $\tv$, and check whether a
valuation exists that satisfies $\phi$ and takes
$\mbox{inline}^{-1}(\tv)$ to ${\cal A}$.\punto
\end{remark}

The vWSDs are already exponentially more succinct than the
v-tabsets. As is easy to verify,

\begin{proposition}
\label{prop:succinctness}
Any v-tabset representation of the WSD
\[
\left\{
\begin{array}{l|c}
C_1 & R.\tid_1.A \\
\hline
& a_1 \\
& b_1
\end{array}
\quad
\cdots
\quad
\begin{array}{l|c}
C_n & R.\tid_n.A \\
\hline
& a_n \\
& b_n
\end{array}
\right\}
\]
where the $a_i, b_i$ are distinct domain values
takes space exponential in $n$.
\end{proposition}

By a similar argument, v(resp.,g)WSDs are exponentially more succinct
than v(g-)TSTs. Succinct attribute-level repersentations have a rather
high price:


\begin{theorem}\label{th:emptyworld}
  Given an attribute-level (g)WSD ${\cal W}$, checking whether the empty
  world is in $rep({\cal W})$ is NP-complete.
\end{theorem}
\begin{proof}
To prove this, we show that the problem is in NP for
attribute-level gWSDs and NP-hard for attribute-level WSDs.

Let ${\cal W} = (\{ C_1, \dots, C_n \}, \phi)$ be a gWSD.  The problem
is in NP since we can nondeterministically check whether there is a
choice of component tuples $\tv_1 \in C_1, \dots, \tv_n \in C_n$ such
that $\tv_1 \circ \dots \circ \tv_n$ represents the empty world.

The proof of NP-hardness is by reduction from Exact Cover by 3-Sets
(X3C) \cite{gj79}. Given a finite set $X$ of size $|X| = 3q$ and a set
$C$ of three-element subsets of $X$, does $C$ contain a subset $C'$
such that every element of $X$ occurs in exactly one member of $C'$?

{\bf Construction.} We construct an attribute-level WSD $\{ C_1,
\dots, C_q \}$ as follows.  Let $C_i$ be a table of schema
$C_i[\tid_1.A_i, \dots, \tid_{|X|}.A_i]$ with tuples $\langle\tid_1.A_i: a_1,
\dots, \tid_{|X|}.A_i: a_{|X|}\rangle$ for each $S \in C$ such that $a_j =
\bot$ if $j \in S$ and $a_j = 1$ otherwise.

{\bf Correctness.}  This is straightforward to show, but note that
each tuple of a component relation contains exactly three $\bot$
symbols. The WSD represents a set of worlds in which each one
contains, naively, up to $3 \cdot q$ tuples.  The composition of $q$
component tuples $w_1 \in C_1, \dots, w_q \in C_q$ can only represent
the empty world if the $\bot$ symbols in $w_1, \dots, w_q$ do not
overlap. This guarantees that $w_1 \circ \dots \circ w_q$ represents
the empty set {\em only if}\/ the sets from $C$ corresponding to $w_1,
\dots, w_q$ form an {\em exact}\/ cover of $X$.\punto
\end{proof}

\begin{example}
We give an example of the previous reduction from X3C to testing whether
the empty world is in the representation of a WSD.
Let $X = \{ 1, 2, 3, 4, 5, 6, 7, 8, 9 \}$ and let
$C = \{ \{ 1,5,9 \}, \{ 2,5,8 \}, \{ 3,4,6 \}, \{ 2,7,8 \}, \{ 1,6,9 \} \}$.
Then the WSD $\{ C_1, C_2, C_3 \}$ with each $C_i$ the table
\[
\begin{small}
\begin{array}{l|ccccccccc}
C_i & \tid_1.A_i & \tid_2.A_i & \tid_3.A_i &
      \tid_4.A_i & \tid_5.A_i & \tid_6.A_i &
      \tid_7.A_i & \tid_8.A_i & \tid_9.A_i \\
\hline
    & \bot & 1 & 1 & 1 & \bot & 1 & 1 & 1 & \bot \\
    & 1 & \bot & 1 & 1 & \bot & 1 & 1 & \bot & 1 \\
    & 1 & 1 & \bot & \bot & 1 & \bot & 1 & 1 & 1 \\
    & 1 & \bot & 1 & 1 & 1 & 1 & \bot & \bot & 1 \\
    & \bot & 1 & 1 & 1 & 1 & \bot & 1 & 1 & \bot \\
\end{array}
\end{small}
\]
for $1 \le i \le 3$ represents the empty world, because every tuple
$\tid_i$ has $\bot$ symbol for some attributes in the result of
combining the first tuple of $C_1$, the third tuple of $C_2$, and the
fourth tuple of $C_3$. Therefore, the first, third and fourth sets in
$C$ are an exact cover of $X$.
\punto
\end{example}

It follows that the problem of deciding whether the $q$-ary tuple
$(1,\dots, 1)$ or whether the world containing just that tuple is {\em
uncertain}\/ is NP-complete.  Note that this NP-hardness is a direct
consequence of the succinctness increase in gWSDs as compared to
gTSTs. On gTSTs, checking for the empty world is a trivial operation.

\begin{corollary}
  Tuple certainty is coNP-hard for attribute-level WSDs.
\end{corollary}

This problem remains in coNP even for general gWSDs.
Nevertheless, since computing certain answers is a central task
related to incomplete information, we will consider also the following
restriction of gWSDs. As we will see, this alternative definition
yields a representation system in which the tuple and instance
certainty problems are in polynomial time while the formalism is still
exponentially more succinct than gTSTs.

\begin{definition}[gWSD]
\em
An attribute-level gWSD is called a {\em tuple-level gWSD}\/
if for any two attributes $A_i, A_j$ from the
schema of relation $R$, and any tuple id $\tid$, the
attributes $R.\tid.A_i, R.\tid.A_j$ of the component tables
are in the same component schema.
\punto
\end{definition}

In other words, in tuple-level gWSDs, values for one and the same
tuple cannot be split across several components -- that is, here the
decomposition is less fine-grained than in attribute-level gWSDs. In the
remainder of this article, we will exclusively study tuple-level (g-,
resp.\ v-)WSDs, and will refer to them as just simply (g-, v-)WSDs.
Obviously, tuple-level (g)WSDs are just as expressive as attribute-level
(g)WSDs, since they all are just decompositions of 1-(g)WSDs.

However, tuple-level (g)WSDs are less succinct than attribute-level
(g)WSDs. For example, any tuple-level WSD equivalent to the attribute-level
WSD
\[
\left\{
\begin{array}{l|c}
C_1 & R.\tid.A_1 \\
\hline
& a_1 \\
& b_1
\end{array}
\quad
\cdots
\quad
\begin{array}{l|c}
C_n & R.\tid.A_n \\
\hline
& a_n \\
& b_n
\end{array}
\right\}
\]
must be exponentially larger.
Note that the WSDs of Proposition~\ref{prop:succinctness} are tuple-level.



\vspace{-3mm}

\section{Main Expressiveness Result}
\label{sec:expr}

In this section we study the expressive power of gWSDs. We show that
gWSDs and c-multitables are equivalent in expressive power, that is,
for each gWSD one can find an equivalent c-multitable that represents
the same set of possible worlds and vice versa. 

\begin{theorem}\label{th:equiv}
gWSDs capture the c-multitables.
\end{theorem}

Thus gWSDs form a strong representation system for relational algebra.

\begin{corollary}\label{cor:gWSDs-strong}
gWSDs are a strong representation system for relational algebra.
\end{corollary}

We prove that gWSDs capture the c-multitables by providing a
translation of gWSDs into x-multitables, a syntactically restricted
form of c-multitables, and a translation of c-multitables into
gWSDs.

\begin{figure}

\vspace{-5mm}

        \begin{center}
        \subfigure[1-gWSD $(\{C_1\},\phi)$]{
        \begin{tabular}{c}
        $\phi = (x \neq 1) \land (x \neq y) \land (z \neq 2)$
        \\
        \\
        \begin{tabular}{c|cccc}
                $C_1$&$R.\tid_1.A$ & $R.\tid_1.B$ & $R.\tid_2.A$ & $R.\tid_2.B$ \\
                \hline
                &$x$ & $y$ & $\bot$ & $\bot$ \\
                &$1$ & $z$ & $z$ & $3$ \\
        \end{tabular}
        \end{tabular}
        \label{fig:ex-wsd-ctab-a}
        }
        \subfigure[x-table ${\cal T}=(T^{\cal T},\phi^{\cal T},\lambda^{\cal T})$]{
        \begin{tabular}{c}
        $\phi^{\cal T} = (x \neq 1) \land (x \neq y) \land (z \neq 2)$
        \\
        \begin{tabular}{c|c@{\extracolsep{0.3cm}}c@{\extracolsep{0.3cm}}l}
                $T^{\cal T}$ & $A$ & $B$ & $cond$ \\
                \hline
                & $x$ & $y$ & $(x_1 = 1)$ \\
                & 1 & $z$ & $(x_1 \neq 1)$ \\
                & $z$ & 3 & $(x_1 \neq 1)$ \\
        \end{tabular}
        \end{tabular}
        \label{fig:ex-wsd-ctab-b}
        }
        \end{center}

        \vspace{-6mm}

        \caption{Translating gWSDs into x-multitables: x-table (b) is equivalent to gWSD (a).}
        \label{fig:ex-wsd-ctab}

        \vspace{-6mm}

        \end{figure}

\begin{lemma}
\label{lem:wsdx-c}
Any gWSD has an equivalent x-multitable of polynomial size.
\end{lemma}

\begin{proof}
  Let ${\cal W} = (\{C_1, \ldots, C_m\}, \phi)$ be a (tuple-level)
  m-gWSD that encodes a g-tabset ${\bf A}$ over relational schema
  $(R_1[U_1], \ldots, R_k[U_k])$.
  
  {\bf Construction.}  We define a translation $f$ from ${\cal W}$ to
  an equivalent c-multitable ${\cal T} = (R_1^{{\cal T}}, \ldots,
  R_k^{{\cal T}}, \phi^{\cal T}, \lambda^{\cal T})$ in the following
  way.

In case a component $C_j$ of ${\cal W}$ is empty, then ${\cal W}$
represents the empty world-set and is equivalent to any x-multitable
with an unsatisfiable global condition, i.e., $x\not=x$. We next
consider the case when all components of ${\cal W}$ are non-empty.
\begin{enumerate}
\item The global condition $\phi$ of ${\cal W}$ becomes the
global condition $\phi^{\cal T}$ of the x-multitable ${\cal T}$.

\item For each relation schema $R_l[U]$ we create a table $R_l^{\cal T}$
with the same schema.

\item We construct a mutex set $\mathbb{M}(\{x_1,\ldots,x_m\},\mu)$
  with $\mu(x_j)=|C_j|-1$ that has a new variable $x_j$ for each
  component $C_j$ of ${\cal W}$. For each local world $w_i\in C_j$
  (with $1\leq i\leq |C_j|$) we create a conjunction
\begin{align*}
  cond(w_i) =
  \begin{cases}
  true & \ldots \textrm{ $\mu(x_j) = 0$}\\
  (x_j = i) & \ldots \textrm{ $1 \leq i \leq \mu(x_j)$}\\
  \overset{\mu(x_j)}{\underset{l=1}{\bigwedge}} (x_j \neq l) & \ldots \textrm{ $i = \mu(x_j)+1$.}
  \end{cases}
\end{align*}
Clearly, any valuation of $x_j$ satisfies precisely one conjunction
$cond(w_i)$. Let $\tid$ be a tuple identifier for a relation $R$
defined in $C_j$, and $\tv$ be the tuple for $\tid$ in $w_i$. If $\tv$
is not a $t_{\bot}$-tuple, then we add $\tv$ with local condition
$\lambda^{\cal T}(\tv)$ to $R_l^{{\cal T}}$, where $R_l^{{\cal T}}$ is
the corresponding table from the x-multitable and $\lambda^{\cal
T}(\tv)$ is the conjunction $cond(w_i)$.
\end{enumerate}

\begin{example}
  Consider the 1-gWSD $(\{C_1\}, \phi)$ given in
  Figure~\ref{fig:ex-wsd-ctab-a}. The first tuple of $C_1$ encodes a
  g-table $R$ with a single tuple (with identifier $\tid_1$), and the
  second tuple of $C_1$ encodes two v-tuples with identifiers $\tid_1$
  and $\tid_2$. The encoding of the gWSD as an x-table ${\cal T}$ with
  global condition $\phi^{\cal T}$ is given in
  Figure~\ref{fig:ex-wsd-ctab-b}. The local conditions of tuples in
  $T^{\cal T}$ are conjunctions from a mutex set
  $\mathbb{M}(\{x_1\},\mu) =\{true,(x_1=1),(x_1\not=1)\}$, where
  $\mu(x_1)=1$. Our translation relies on the fact that any valuation
  of the mutex variables satisfies precisely one (in)equality for each
  mutex variable. For instance, if the first tuple of $T^{\cal T}$
  would have the local condition $x_1=2$, then a valuation
  $\{x_1\mapsto 2\}$ would wrongly allow worlds containing
  instantiations of the first two tuples of $T^{\cal T}$, although
  this is forbidden by our gWSD.\punto
\end{example}

{\bf Correctness.}  Take the g-tabset ${\bf A}$ represented by ${\cal
W}$: $${\bf A} = \{(\mbox{inline}^{-1}(w_1\times \ldots \times w_m),\phi)
\mid \underset{j=1}{\overset{m}{\bigwedge}} (w_j\in C_j)\}.$$ 

We create a g-tabset ${\bf A}'$ that consists of the g-multitables of
${\bf A}$ with global conditions $\phi$ enriched by conjunctions from
our mutex set $\mathbb{M}$ such that precisely one of these
conjunctions is true for any valuation of the mutex variables. We
consider then a new global condition $\phi_{(w_1,\ldots,w_m)}
:=\phi\wedge cond(w_1)\wedge\ldots\wedge cond(w_m)$ for each
g-multitable ${\cal B}_{(w_1,\ldots,w_m)}$ defined by
$\mbox{inline}^{-1}(w_1\times\ldots
\times w_m)$ with initial global condition $\phi$. 

Clearly, ${\bf A}'$ is equivalent to ${\bf A}$, because there is a
ono-to-one mapping between g-multitables of ${\bf A}$ and of ${\bf
A}'$, respectively. A choice of a g-multitable from ${\bf A}$, or any
world ${\cal A}$ it represents, is then precisely mapped to its
corresponding g-multitable from ${\bf A}'$, or world ${\cal A}$, under
an appropriate assignment of the mutex variables. This also holds for
the other direction. 

We next show that $rep({\bf A}') = rep ({\cal T})$.

Any total valuation $\nu$ over the mutex variables $x_1,\ldots,x_m$ is
identity on $\phi$ and satisfies precisely one conjunction
$cond(w_1)\wedge\ldots\wedge cond(w_m)$: $$\nu({\bf A}') =
(\{(\nu({\cal B}_{(v_1,\ldots,v_m)}),\nu(\phi_{(v_1,\ldots,v_m)}))\mid
1\leq j\leq m, v_j\in C_j\}) = ({\cal B}_{(w_1,\ldots,w_m)},\phi).$$

Let ${\cal B} = ({\cal B}_{(w_1,\ldots,w_m)},\phi)$ for short. It
remains to show that $rep({\cal B}) = rep(\nu({\cal T}))$.

{\noindent ($\subseteq$)} The translation $f$ maps each tuple of a
table $R_l^{\cal B}$ to an identical tuple in $R_l^{\cal T}$, where
$R_l\in\{R_1,\ldots,R_k\}$. We also have $\nu(\phi)=\phi=\phi^{\cal
T}$. Thus $R_l^{\cal B}\subseteq R_l^{\cal T}$ in each world
represented by ${\cal T}$.

{\noindent ($\supseteq$)} Assume there is a tuple $\tv\in\nu(R_l^{\cal
T})$ such that $\tv\not\in R_l^{\cal B}$. The translation $f$ ensures
that $\tv$ comes from a combination of local worlds
$(c_1,\ldots,c_m)$, which corresponds to a g-multitable ${\cal B}'$
with global condition $\phi\wedge cond(c_1)\wedge\ldots\wedge
cond(c_m)$. We thus have that $\nu(cond(c_1)\wedge\ldots\wedge
cond(c_m)) = true$ for $\tv$ to be defined by ${\cal B}'$. However,
there is only one combination of local worlds with this property,
namely $(w_1,\ldots,w_m)$, which defines ${\cal B}$. Contradiction.

{\bf Complexity.} By construction, the translation $f$ is the identity
for global conditions and maps each tuple $\tv$ defined by a component
of ${\cal W}$ and different from $t_{\bot}$ to precisely one tuple of
of a table of ${\cal T}$ with local condition of polynomial size. The
translation $f$ is thus polynomial. \punto
\end{proof}

\begin{figure}
\begin{center}
\subfigure[c-table ${\cal T}$]{
	 \begin{tabular}{c}
                $\phi^{\cal T} = (x \neq 1) \land (x = z)$\\
		\\
                \begin{tabular}{c|c@{\extracolsep{0.4cm}}c@{\extracolsep{0.5cm}}l}
                $T^{\cal T}$ & $A$ & $B$ & $cond$ \\
                \hline
                $\tid_1$ & $x$ & 1 & $(x \neq 2)$ \\
                $\tid_2$ & $z$ & $y$ & $(y \neq 2)$ \\	
	        \end{tabular}
	        \\
	        \\
	 \end{tabular}
        \label{fig:ex-ctab-wsd-a}\vspace*{-2em}
}
\hspace{1cm}
\subfigure[An optimized 1-gWSD equivalent to c-table ${\cal T}$.]{
	 \begin{tabular}{c}
                $\phi = (x \neq 1) \wedge (x \neq 2) \wedge (y \neq 2)$\\
		\\
                \begin{tabular}{l|cccc}
                $C$ & $R.\tid_1.A$ & $R.\tid_1.B$ & $R.\tid_2.A$ & $R.\tid_2.B$ \\
                \hline
		& $\bot$ & $\bot$ & $\bot$ & $\bot$ \\
		& $\bot$ & $\bot$ & 2 & $y$ \\
		& $x$    & 1      & $x$ & $y$ \\
		& $x$    & 1      & $\bot$ & $\bot$ \\
	        \end{tabular}
	        \\
	 \end{tabular}
        \label{fig:ex-ctab-wsd-b}\vspace*{-2em}
}

\subfigure[1-gWSD equivalent to c-table ${\cal T}$. The $\Theta$'s are given here for clarity.]{
        \begin{tabular}{c}
        $\phi = (x \neq 1) \wedge (x \neq 2) \wedge (y \neq 1) \wedge (y \neq 2) \wedge (x \neq y)$\\
        \\
        \begin{tabular}{l|cccc}
                $C$ & $R.\tid_1.A$ & $R.\tid_1.B$ & $R.\tid_2.A$ & $R.\tid_2.B$ \\
                \hline
                 $\Theta_1 :=  (x \not= 1 \wedge x = z \wedge x = 2 \wedge x = y)$  
	& $\bot$ & $\bot$ & $\bot$ & $\bot$ \\
		 $\Theta_2 :=  (x \not= 1 \wedge x = z \wedge x = 2 \wedge x \not= y \wedge y \not= 2)$ 
	& $\bot$ & $\bot$ & 2 & $y$ \\
		 $\Theta_3 :=  (x \not= 1 \wedge x = z \wedge x = 2 \wedge x \not= y \wedge y = 1)$ 
	& $\bot$ & $\bot$ & 2 & 1 \\
		 $\Theta_4 :=  (x \not= 1 \wedge x = z \wedge x = 2 \wedge x \not= y \wedge y \not= 1)$
	& $\bot$ & $\bot$ & 2 & $y$ \\
		 $\Theta_5 :=  (x \not= 1 \wedge x = z \wedge x \not= 2 \wedge x = y)$
	& $x$ & $1$ & $x$ & $x$ \\
		 $\Theta_6 :=  (x \not= 1 \wedge x = z \wedge x \not= 2 \wedge x \not= y \wedge y = 1)$
	& $x$ & $1$ & $x$ & 1 \\
		 $\Theta_7 :=  (x \not= 1 \wedge x = z \wedge x \not= 2 \wedge x \not= y \wedge y\not=1)$
	& $x$ & $1$ & $x$ & $y$ \\
		 $\Theta_8 :=  (x \not= 1 \wedge x = z \wedge x \not= 2 \wedge x \not= y \wedge y = 2)$
	& $x$ & $1$ & $\bot$ & $\bot$ \\
		 $\Theta_9 :=  (x \not= 1 \wedge x = z \wedge x \not= 2 \wedge x \not= y \wedge y \not= 2)$
	& $x$ & $1$ & $x$ & $y$ \\
        \end{tabular}
	\\
	\\
        \end{tabular}
        \label{fig:ex-ctab-wsd-c}
}
\end{center}\vspace*{-1em}

\caption{Translating c-tables into 1-gWSDs.}
\label{fig:ex-ctab-wsd}
\vspace*{-1em}
\end{figure}

For the other, somewhat more involved direction, we first show that
c-multitables can be translated into equivalent g-tabsets.  That is,
disjunction on the level of entire tables plus conjunctions of negated
equalities as global conditions, as present in g-tables, are enough to
capture the full expressive power of c-tables. In particular, we are
able to eliminate all local conditions.

\begin{proposition}
\label{lem:c-gtabset}
Any c-multitable has an equivalent g-tabset.
\end{proposition}
\begin{proof}
Let ${\cal T} = (R_1^{\cal T},\ldots,R_k^{\cal T}, \phi^{\cal T},
\lambda^{\cal T})$ be a c-multitable over relational schema $(R_1[U_1]$, $\ldots$, 
$R_k[U_k])$; $\phi^{\cal T}$ is the global condition and
$\lambda^{\cal T}$ maps each tuple to its local condition. Let ${\bf
X}_{\cal T}$ and ${\bf D}_{\cal T}$ be the set of all variables and
the set of all constants appearing in the c-multitable, respectively.

{\bf Construction.}  We construct a g-tabset ${\bf G}$ with
g-multitables over the same schema $(R_1[U_1],\ldots,R_k[U_k])$ as
follows. We consider comparisons of the form $\tau = \tau'$ and $\tau
\neq \tau'$ where $\tau, \tau' \in {\bf X}_{\cal T} \cup {\bf D}_{\cal T}$ are
variables or constants from the c-multitable.  We compute the set of
all consistent $\Theta = \bigwedge \{ \tau \,
\theta_{\tau,\tau'} \,
\tau'\mid \tau, \tau' \in {\bf X}_{\cal T} \cup {\bf D}_{\cal T} \} $ where
$\theta_{\tau,\tau'} \in \{ =, \neq \}$ for all $\tau$, $\tau'$ and
$\Theta \vDash \phi^{\cal T}$. Note that the equalities in $\Theta$
define an equivalence relation on ${\bf X}_{\cal T} \cup {\bf D}_{\cal
T}$.  In particular, we take into account that $c = c'$ is consistent
iff $c$ and $c'$ are the same constant. We denote by $[x_i]_=$ the
equivalence class of a variable $x_i$ with respect to the equalities
given by $\Theta$ and by $h([x_i]_=)$ the representative element of
that equivalent class (e.g.\ the first element with respect to any
fixed order of the elements in the class).

For each $\Theta$, we construct a g-multitable ${\cal G}_{\Theta}$ in
${\bf G}$. Each tuple $t$ from a table $R_i^{\cal T}$ becomes a tuple
in $R_i^{{\cal G}_{\Theta}}$ if $\Theta\vDash \lambda^{\cal
T}(t)$. The global condition of ${\cal G}_{\Theta}$ is $\Theta$. To
strictly adhere to the definition of g-multitables, we remove the
equalities from $\Theta$ and enforce them in the tables $R_1^{{\cal
G}_{\Theta}},\ldots,R_k^{{\cal G}_{\Theta}}$: In case of a tuple
$\tuple{x_1,\ldots,x_n}$, we replace $x_i$ by $c$ in case $c \in {\bf
D}_{\cal T}, \Theta \vDash (x_i = c)$, and by $h([x_i])$ in case
$\forall c \in {\bf D}_{\cal T}, \Theta \vDash (x_i \neq c)$.

{\bf Correctness.} Clearly, the g-tabset ${\bf G}$ consists of a
finite number of g-multitables, because the finite number of variables
and constants in ${\cal T}$ induces finitely many consistent
$\Theta$'s. We next show that $rep({\bf G}) = rep({\cal T})$.

($\subseteq$) Given a world ${\cal A}$ represented by a g-multitable
${\cal G}_{\Theta}\subseteq {\bf G}$ for a conjunction $\Theta$. For
simplicity, we consider the (equivalent) multitable where the
equalities are not removed from $\Theta$ and also not propagated in
the g-tables. By construction, $\Theta\vDash \phi^{\cal T}$ and a
tuple $t$ is in a table $R^{{\cal G}_{\Theta}}$ if it occurs in a
table $R^{\cal T}$ such that $\Theta\vDash \lambda^{\cal T}(t)$. Thus
we necessarily have that ${\cal A}\in rep({\cal T})$.

($\supseteq$) Given a world ${\cal A}\in rep(\nu({\cal T}))$ defined
by a total valuation $\nu$ consistent with $\phi^{\cal T}$. Because
$\nu$ and $\Theta$ talk about the same set of variables and there is a
$\Theta$ for each possible (in)equality on any two variables or
variable and constants that are consistent with $\phi^{\cal T}$, there
exists a consistent $\Theta$ such that $\Theta\vDash\nu$. Let ${\cal
G}_{\Theta}$ be the g-multitable in ${\bf G}$ for our chosen
$\Theta$. Take now any tuple $t$ in a table $R^{\cal T}$ such that
$\nu(\lambda^{\cal T}(t))=true$. Then, because $\Theta\vDash\nu$ we
have $\Theta\vDash\lambda^{\cal T}(t)$ and $t\in R^{{\cal
G}_{\Theta}}$. Thus ${\cal A}\in rep({\cal G}_{\Theta})\subseteq
rep({\bf G})$. \punto
\end{proof}

Any g-tabset can be inlined into a g-TST, which, by the definition of
gWSDs, represents a 1-gWSD. It then follows that

\begin{lemma}
\label{lem:c-wsdx}
Any c-multitable has an equivalent gWSD.
\end{lemma}

\begin{example}
Figure~\ref{fig:ex-ctab-wsd-a} shows a c-table ${\cal T}$. Following
the construction from the proof of Proposition~\ref{lem:c-gtabset}, we
create nine consistent $\Theta$'s and one g-table for each of them.
Figure~\ref{fig:ex-ctab-wsd-c} shows the $\Theta$'s and an inlining of
all these g-tables into a gTST. The gTST is normalized by creating one
common global condition. This gTST with a global condition of
inequalities is in fact a 1-gWSD. Figure~\ref{fig:ex-ctab-wsd-b} shows
a simplified version of our 1-gWSD, where duplicate tuples are removed
and some different tuples are merged. For instance, the tuple for
$\Theta_4$ is equal to the tuple for $\Theta_1$ and can be
removed. Also, by merging the tuples for $\Theta_2$ and $\Theta_3$ we
also allow $y$ to take value 1 and thus we eliminate the inequality
$y\not=1$ form the global condition $\phi$.\punto
\end{example}

As a corollary of Lemma~\ref{lem:wsdx-c}, x-multitables, a
syntactically restricted form of c-multitables, are at least as
expressive as gWSDs. However, by Lemma~\ref{lem:c-wsdx}, gWSDs are at
least as expressive as c-multitables. This implies that

\begin{corollary}
  The x-multitables capture gWSDs and thus c-multitables.
\end{corollary}

To sum up, we can chart the expressive power of the representation
systems considered in this paper as follows. As discussed in
Section~\ref{sec:wsd}, v-multitables are less expressive than finite
sets of v-multitables (or v-tabsets), which are syntactic variations
of vTSTs.  The vWSDs (resp., gWSDs) are equally expressive to v(g)TSTs
yet exponentially more succinct
(Proposition~\ref{prop:succinctness}). The gWSDs are more expressive
than vWSDs because gWSDs can represent the answers to any relational
algebra query, whereas vWSDs cannot represent answers to queries with
selections or difference. Finally, c-multitables are captured by their
syntactic restriction called x-multitables and also by gWSDs.



\vspace{-3mm}

\section{Complexity of Managing gWSDs}
\label{sec:decision}

\begin{figure}[t]
  \centering

  \vspace{-5mm}

        \begin{center}
        \subfigure[3-gWSD ${\cal W}=(\{C_1,C_2,C_3\},\mathit{true})$]{
          \begin{tabular}{c|ccc}
                $C_1$ & $R.\tid_1.A$ & $R.\tid_2.A$ & $S.\tid_1.B$ \\
                \hline
                      & $2$          & $y$          & $z$ \\
                      & $\bot$       & $2$          & $2$ \\
        \end{tabular}%
        \hspace*{1em}%
        \begin{tabular}{c|c}
                $C_2$ & $R.\tid_3.A$ \\
                \hline
                      & $1$\\
        \end{tabular}%
        \hspace*{1em}%
        \begin{tabular}{c|c}
                $C_3$ & $S.\tid_2.B$ \\
                \hline
                      & $1$\\
                      & $2$\\
        \end{tabular}
        }%
        \subfigure[x-multitable ${\cal T}$ with $\phi^{\cal T}=\textit{true}$]{
                
          \begin{tabular}{c|c@{\extracolsep{0.3cm}}l@{\extracolsep{0.3cm}}l}
            $R^{\cal T}$ & $A$ & $cond$ \\
            \hline
            & $2$ & $x_1 = 1$ \\
            & $y$ & $x_1 = 1$ \\
            & $2$ & $x_1 \neq 1$ \\
            & $1$ & $\mathit{true}$ \\
          \end{tabular}%
          \hspace*{1em}%
          \begin{tabular}{c|c@{\extracolsep{0.3cm}}l@{\extracolsep{0.3cm}}l}
            $S^{\cal T}$ & $B$ & $cond$ \\
            \hline
            & $z$ & $x_1 = 1$ \\
            & $2$ & $x_1 \neq 1$ \\
            & $1$ & $x_3 = 1$ \\
            & $2$ & $x_3 \neq 1$ \\
          \end{tabular}
        }
        \end{center}

        \vspace{-8mm}

  \caption{Example of a 3-gWSD and an equivalent x-multitable.}
  \label{fig:decision-tuple}
\end{figure}

We consider the data complexity of the decision problems defined in
Section~\ref{sec:introduction}. Note that in the literature the tuple
(q-)possibility and (q-)certainty problems are sometimes called
bounded or restricted (q-)possibility, and (q-)certainty respectively,
and the instance (q-)possibility and (q-)certainty are sometimes
called (q-)membership and (q-)uniqueness~\cite{AKG1991}. A comparison
of the complexity results for these decision problems in the context
of gWSDs to those of c-tables~\cite{AKG1991} and Trio~\cite{trio} is
given in Table~\ref{tab:decision-overview}.
%

\subsection{Tuple (q)-possibility}

We first prove complexity results for tuple q-possibility in the
context of x-tables. This is particularly relevant as gWSDs can be
translated in polynomial time into x-tables, as done in the proof of
Lemma~\ref{lem:wsdx-c}.

\begin{lemma}
        \label{lemma:xqposs}
        Tuple q-possibility is in PTIME for x-tables and positive
        relational algebra.
\begin{proof}\em
  Recall from Definition~\ref{def:xtables} and
  Proposition~\ref{prop:xtables} that x-tables are closed under
  positive relational algebra and the evaluation of positive
  relational algebra queries on x-tables is in PTIME.
  
  Consider a constant tuple $t$ and a fixed positive relational query $Q$,
  both over schema $U$, and two x-multitables ${\cal T}$ and ${\cal T}'$ such
  that ${\cal T}' = Q({\cal T})$.
  
  In case the global condition of ${\cal T}'$ is unsatisfiable, then ${\cal
    T}'$ represents the empty world-set and $t$ is not possible.  The global
  condition is a conjunction of negated equalities and we can check its
  unsatisfiability in PTIME. We consider next the case of satisfiable global
  conditions. Following the semantics of x-tables, the tuple $t$ is possible
  in ${\cal T}'$ iff there is a tuple $t'$ in ${\cal T}'$ and a valuation
  $\nu$ consistent with the global and local conditions such that $t'$ equals
  $t$ under $\nu$.  This can be checked for each ${\cal T}'$-tuple
  individually and in PTIME. \punto
\end{proof}
\end{lemma}

\begin{theorem}
        \label{th:t-qposs}
        Tuple q-possibility is in PTIME for gWSDs and positive
        relational algebra.
\begin{proof}\em
  This follows from the polynomial time translation of gWSDs into
  x-multitables ensured by Lemma~\ref{lem:wsdx-c} and the PTIME result
  for x-multitables given in Lemma~\ref{lemma:xqposs}.\punto
\end{proof}
\end{theorem}

For full relational algebra, tuple q-possibility becomes NP-hard even
for v-tables where each variable occurs at most once (also called Codd
tables)~\cite{AKG1991}.

\begin{theorem}
        \label{th:t-qposs-all}
        Tuple q-possibility is in NP for gWSDs and relational algebra
        and NP-hard for WSDs and relational algebra.
\begin{proof}\em
  Tuple q-possibility is in NP for gWSDs and relational algebra
  because gWSDs can be translated polynomially into c-tables (see
  Lemma~\ref{lem:wsdx-c}) and tuple q-possibility is in NP for
  c-tables and relational algebra~\cite{AKG1991}.

  We show NP-hardness for WSDs and relational algebra by a reduction
  from 3CNF-satisfiability~\cite{gj79}. Given a set \textbf{Y} of
  propositional variables and a set of clauses $c_i=c_{i,1}\vee
  c_{i,2}\vee c_{i,3}$ such that for each $i,k$, $c_{i,k}$ is $x$ or
  $\neg x$ for some $x\in\mathbf{Y}$, the 3CNF-satisfiability problem
  is to decide whether there is a satisfying truth assignment for
  $\bigwedge_i c_i$.

  \textbf{Construction.} We create a WSD ${\cal
    W}=(C_1,\ldots,C_{|\textbf{Y}|},C_S)$ representing worlds of two
  relations $R$ and $S$ over schemas $R(C)$ and $S(C)$, respectively,
  as follows\footnote{For clarity reasons, we use two relations; they
    can be represented as one relation with an additional attribute
    stating the relation name.}. For each variable $x_i$ in \textbf{Y}
  we create a component $C_i$ with two local worlds, one for $x_i$ and
  the other for $\neg x_i$.  For each literal $c_{i,k}$ we create an
  $R$-tuple $\langle i\rangle$ with id $\tid_{i,k}$. In case $c_{i,k}
  = x_j$ or $c_{i,k} = \neg x_j$, then the schema of $C_j$ contains
  the attribute $R.\tid_{i,k}.C$ and the local world for $x_j$ or
  $\neg x_j$, respectively, contains the values $\langle i\rangle$ for
  these attributes. All component fields that remained unfilled are
  finally filled in with $\bot$-values. The additional component $C_S$
  has $n$ attributes $S.\tid_1.C,\ldots,S.\tid_n.C$ and one local
  world $(1,\ldots,n)$. Thus, by construction, $S=\{ \langle
  C:1\rangle,\ldots,\langle C:n\rangle\}$ and $R\subseteq S$ in all
  worlds defined by ${\cal W}$.

  The problem of deciding whether $\bigwedge_i c_i$ has a satisfying
  truth assignment is equivalent to deciding whether the nullary tuple
  $\tuple{}$ is possible in the answer to the fixed query $Q
  =\{\tuple{}\}-\pi_{\emptyset}(S- R)$, with $S$ and $R$ defined by
  ${\cal W}$.

  \textbf{Correctness.}  Clearly, $\tuple{}$ is possible in the answer
  to $Q$ iff there is a world ${\cal A}\in \mathit{rep}({\cal W})$
  where $\pi_{\emptyset}(S- R)$ is empty, or equivalently $S-R$ is
  empty. Because by construction $R\subseteq S$ in all worlds defined
  by ${\cal W}$, we further refine our condition to $\exists {\cal
    A}\in \mathit{rep}({\cal W}): S^{\cal A}=R^{\cal A}$.  We next
  show that $\bigwedge_i c_i$ has a satisfying truth assignment
  exactly when $\exists {\cal A}\in \mathit{rep}({\cal W}): S^{\cal
    A}=R^{\cal A}$.
  
  First, assume there is a truth assignment $\nu$ of \textbf{Y} that
  proves $\bigwedge_i c_i$ is satisfiable. Then, $\nu(c_i)$ is
  \textit{true} for each clause $c_i$.  Because each clause $c_i$ is a
  disjunction, this means there is at least one $c_{i,k}$ for each
  $c_i$ such that $\nu(c_{i,k})$ is \textit{true}.
  
  Turning to ${\cal W}$, $\nu$ represents a choice of local worlds of
  ${\cal W}$ such that for each variable $x_j\in\mathbf{Y}$ if
  $\nu(x_j)=\mathit{true}$ then we choose the first local world of
  $C_j$ and if $\nu(x_j)=\mathit{false}$ then we choose the second
  local world of $C_j$.  Let $w_j$ be the choice for $C_j$ and let
  $w_{C_S}$ be the only choice for $C_S$. Then, ${\cal W}$ defines a
  world ${\cal A} =\mbox{inline}^{-1}(w_1\times\ldots\times
  w_{|\textbf{Y}|}\times w_{C_S})$ and $R^{\cal A}$ contains those
  tuples defined in the chosen local worlds.  Because there is at
  least one $c_{i,k}$ per clause $c_i$ such that $\nu(c_{i,k})$ is
  \textit{true}, there is also a local world $w_j$ that defines
  $R$-tuple $\langle C:i\rangle$ for each $c_i$. Thus $R^{\cal
    A}=S^{\cal A}$.
        
  Now, assume there exists a world ${\cal A}\in\mathit{rep}({\cal W})$
  such that $S^{\cal A}=R^{\cal A}$. Thus $R^{\cal A}=\{\langle
  C:1\rangle,\ldots,\langle C:n\rangle\}$ and there is a choice of
  local worlds of the components in ${\cal W}$ that define all
  $R$-tuples $\langle C:1\rangle$ through $\langle C:n\rangle$. By
  construction, this choice corresponds to a truth assignment $\nu$
  that maps at least one literal $c_{i,k}$ of each $c_i$ to
  \textit{true}. Thus $\nu$ is a satisfying truth assignment of
  $\bigwedge_i c_i$.  \punto
\end{proof}
\end{theorem}

The construction used in the proof of Theorem~\ref{th:t-qposs-all} can
be also used to show that instance possibility is NP-hard for
(tuple-level) WSDs: deciding the satisfiability of 3CNF is reducible
to deciding whether the relation
$\{\tuple{C:1},\dots,\tuple{C:n}\}$ is a possible instance of
$R$.

\begin{figure}[t!]
\begin{center}
  \begin{align*}
    \mbox{3CNF clause set: } \{c_1 = x_1\vee x_2\vee x_3,\hspace*{2em}
    c_2 = x_1\vee \neg x_2\vee x_4, \hspace*{2em}
    c_3 = \neg x_1\vee x_2\vee \neg x_4\}
  \end{align*}
  \hspace{0.4cm}
  \begin{tabular}{c|c@{\hspace*{.5em}}c@{\hspace*{.5em}}c}
    $C_1$ & $R.\tid_{1,1}$.C & $R.\tid_{2,1}$.C & $R.\tid_{3,1}$.C \\
    \hline
    $(x_1)$      & 1 & 2 & $\bot$ \\
    $(\neg x_1)$ & $\bot$ & $\bot$ & 3 \\
  \end{tabular}
  \hspace{0.4cm}
  \begin{tabular}{c|c@{\hspace*{.5em}}c@{\hspace*{.5em}}c}
    $C_2$ & $R.\tid_{1,2}$.C & $R.\tid_{2,2}$.C & $R.\tid_{3,2}$.C \\
    \hline
    $(x_2)$      & 1 & $\bot$  & 3 \\
    $(\neg x_2)$ & $\bot$ & 2 & $\bot$  \\
  \end{tabular}
  \hspace{1cm}\vspace*{1em}

  \begin{tabular}{c|c}
    $C_3$ & $R.\tid_{1,3}$.C \\
    \hline
    $(x_3)$      & 1 \\
    $(\neg x_3)$ & $\bot$  \\
  \end{tabular}
  \hspace{0.4cm}
  \begin{tabular}{c|c@{\hspace*{.5em}}c}
    $C_4$ & $R.\tid_{2,3}$.C & $R.\tid_{3,3}$.C \\
    \hline
    $(x_4)$      & 2 & $\bot$ \\
    $(\neg x_4)$ & $\bot$ & 3  \\
  \end{tabular}
  \hspace{0.4cm}
  \begin{tabular}{c|c@{\hspace*{.5em}}c@{\hspace*{.5em}}c}
    $C_S$ & $S.\tid_1$.C & $S.\tid_2$.C & $S.\tid_3$.C \\
    \hline
    $w_{C_S}$      & 1 & 2 & 3 \\
  \end{tabular}
\end{center}
\caption{3CNF clause set encoded as WSD.}
\label{fig:tuple-qposs}
\end{figure}

\begin{example} 
  Figure~\ref{fig:tuple-qposs} gives a 3CNF clause set and its WSD
  encoding. Checking the satisfiability of $c_1\wedge c_2\wedge c_3$
  is equivalent to checking whether there is a choice of local worlds
  in the WSD such that $\tuple{}$ is possible in the answer to the
  query $\{\tuple{}\}-\pi_{\emptyset}(S- R)$, or, simpler, such that
  $S-R$ is empty. This also means that $R = \{\langle C:1\rangle,
  \langle C:2\rangle, \langle C:3\rangle\}$.  For example,
  $\{x_1\mapsto \mathit{true}, x_2\mapsto \mathit{true}, x_3\mapsto
  \mathit{true},x_4\mapsto \mathit{true}\}$ is a satisfying truth
  assignment. Indeed, the corresponding choice of local worlds
  $(C_1:x_1,C_2:x_2,C_3:x_3,C_4:x_4,C_S:w_{C_S})$ defines a world
  ${\cal A}$ in which $R^{\cal A} = S^{\cal A}$.\punto
\end{example}

The result for tuple possibility follows directly from
Theorem~\ref{th:t-qposs}, where the positive relational query is the
identity.

\begin{theorem}
  \label{th:t-poss}
  Tuple possibility is in PTIME for gWSDs.
\end{theorem}

Recall from Table~\ref{tab:decision-overview} that tuple possibility
is NP-complete for c-tables. This is because deciding whether a tuple
is possible requires to check satisfiability of local conditions,
which can be arbitrary Boolean formulas.

\subsection{Instance (q)-possibility}

\begin{theorem}
        \label{th:i-poss} Instance possibility is in NP for gWSDs and
        NP-hard for WSDs.
\end{theorem}
\begin{proof}
Let ${\cal W} = (\{ C_1, \dots, C_n \}, \phi)$ be a gWSD. The problem is in
NP since we can nondeterministically check whether there is a choice
of tuples $\tv_1 \in C_1, \dots, \tv_n \in C_n$ such that $\tv_1
\circ\dots \circ \tv_n$ represents the input instance.

We show NP-hardness for WSDs with a reduction from {\textsf Exact
Cover by 3-Sets} \cite{gj79}.
                
Given a set $X$ with $|X| = 3q$ and a collection $C$ of 3-element
subsets of $X$, the exact cover by 3-sets problem is to decide whether
there exists a subset $C' \subseteq C$, such that every element of $X$
occurs in exactly one member of $C'$.
                                                
    {\bf Construction.} The set $X$ is encoded as an instance
    consisting of a unary relation $I_X$ over schema $I_X[A]$ with
    $3q$ tuples. The collection $C$ is represented as a WSD ${\cal W} =
    \{C_1, \ldots, C_q\}$ encoding a relation $R$ over schema $R[A]$,
    where $C_1, \ldots, C_q$ are component relations.  The schema of a
    component $C_i$ is\\ $C_i[R.\tid_{j + 1}.A, R.\tid_{j + 2}.A,
    R.\tid_{j + 3}.A]$, where $j = \lfloor \frac{i}{3} \rfloor$. Each
    3-element set $c = \{x,y,z\}$ $\in C$ is encoded as a tuple
    $(x,y,z)$ in each of the components $C_i$.
                                                
    The problem of deciding whether there is an exact cover by 3-sets
    of $X$ is equivalent to deciding whether $I_X \in rep({\cal W})$.

{\bf Correctness.} We prove the correctness of the reduction, that is,
we show that $X$ has an exact cover by 3-sets exactly when $I_X \in
rep({\cal W})$.

First, assume there is a world ${\cal A} \in rep({\cal W})$ with $R^{\cal A}
= I_X$. Then there exist tuples $w_i \in C_i, 1 \leq i \leq q$, such
that ${\cal A} = rep(\{w_1\} \times \ldots \times \{w_q\})$. As $I_X$
and $R^{\cal A}$ have the same number of tuples and all elements of
$I_X$ are different, it follows that the values in $w_1, \ldots, w_q$
are disjoint. But then this means that the elements in $w_1, \ldots,
w_q$ are an exact cover of $X$.
        
Now, assume there exists an exact cover $C'=\{c_1, \ldots, c_q\}$ of
$X$.  Let $w_i \in C_i$ such that $w_i = c_i, 1 \leq i \leq q$. As the
elements $c_i$ are disjoint, the world ${\cal A} = rep(\{w_1\} \times
\ldots \times \{w_q\})$ contains exactly $3q$ tuples. Since $C'$ is an
exact cover of $X$ and each element of $X$ (and therefore of $I_X$)
appears in exactly one local world $w_i$, it follows that $I_X =
R^{\cal A}$.
\punto
\end{proof}

\begin{figure}[t!]
\begin{center}
        \begin{tabular}{c|c}
                $I_X$ & A \\
                \hline
                & 1 \\
                & 2 \\
                & 3 \\
                & 4 \\
                & 5 \\
                & 6 \\
                & 7 \\
                & 8 \\
                & 9 \\
        \end{tabular}
        \hspace{0.4cm}
        \begin{tabular}{c|ccc}
        $C_1$ & $t_1$.A & $t_2$.A & $t_3$.A \\
                \hline
                $w_1$ & 1 & 5 & 9 \\
                $w_2$ & 2 & 5 & 8 \\
                $w_3$ & 3 & 4 & 6 \\
                $w_4$ & 2 & 7 & 8 \\
                $w_5$ & 1 & 6 & 9 \\
        \end{tabular}
        \hspace{0.4cm}
        \begin{tabular}{c|ccc}
                $C_2$ & $t_4$.A & $t_5$.A & $t_6$.A \\
                \hline
                $w_1$ & 1 & 5 & 9 \\
                $w_2$ & 2 & 5 & 8 \\
                $w_3$ & 3 & 4 & 6 \\
                $w_4$ & 2 & 7 & 8 \\
                $w_5$ & 1 & 6 & 9 \\
        \end{tabular}
        \hspace{0.4cm}
        \begin{tabular}{c|ccc}
                $C_3$ & $t_7$.A & $t_8$.A & $t_9$.A \\
                \hline
                $w_1$ & 1 & 5 & 9 \\
                $w_2$ & 2 & 5 & 8 \\
                $w_3$ & 3 & 4 & 6 \\
                $w_4$ & 2 & 7 & 8 \\
                $w_5$ & 1 & 6 & 9 \\
        \end{tabular}
\end{center}
\caption{Exact cover by 3-sets encoded as WSD.}
\label{fig:inst-poss}
\end{figure}

\begin{example} 
        Consider the set $X$ and the collection of 3-element sets $C$ defined as
        \begin{eqnarray}
                X &=& \{1,2,3,4,5,6,7,8,9\}\nonumber\\
                C &=& \{\{1,5,9\}, \{2,5,8\}, \{3,4,6\}, \{2,7,8\}, \{1,6,9\}\}\nonumber
        \end{eqnarray}

        The encoding of $X$ and $C$ is given in
        Figure~\ref{fig:inst-poss} as WSD ${\cal W}$ and instance $I_X$. A
        possible cover of $X$, or equivalently, a world of $rep({\cal W})$
        equivalent to $I_X$, is the world $\mbox{inline}^{-1}(w_1\circ
        w_3\circ w_4)$ or, by resolving the record composition,
        \begin{align*}
          \mbox{inline}^{-1}(t_1.A:1,t_2.A:5,t_3.A:9,t_4.A:3,t_5.A:4,t_6:A:6,t_7:2,t_8:7,t_9.A:8).
      \end{align*}\punto
\end{example}
\nop{
  \begin{proof}[Sketch] \em A proof of this result is presented in
  \cite{AKO06gWSD-Full}; here we only give a brief sketch. Checking membership
  in
  NP is straightforward once we have guessed which tuples from the
  gWSD components constitute the input  instance. NP-hardness can be
  shown by reduction from Exact Cover by 3-Sets. Given a
  set $X$ with $|X| = 3q$ and a set $C$ of 3-subsets of $X$, we turn
  $X$ into a unary relation and $C$ into a ternary relation. We
  construct a WSD by taking $q$ copies of $C$ as component
  relations. There is an exact cover of $X$ by 3-sets from $C$ iff $X$
  is a world in the WSD representation. \punto \end{proof}
}

\begin{theorem}
  \label{th:i-qposs}
  Instance q-possibility is NP-complete for gWSDs and relational
  algebra.
  \begin{proof} \em For the identity query, the problem becomes
    instance possibility, which is NP-complete (see
    Theorem~\ref{th:i-poss}).  To show it is in NP, we use the PTIME
    reduction from gWSDs to c-tables given in Lemma~\ref{lem:wsdx-c}
    and the NP-completeness result for instance q-possibility and
    c-tables~\cite{AKG1991}.\punto
  \end{proof}
\end{theorem}

\subsection{Tuple and instance certainty}

\begin{theorem}
  \label{th:t-cert} 
Tuple certainty is in PTIME for gWSDs.

\begin{proof}\em
Consider a tuple-level gWSD $W = (\{C_1,\ldots, C_m\},\phi)$ and a
tuple $t$. Tuple $t$ is certain exactly if $\phi$ is unsatisfiable or
there is a component $C_i$ such that each tuple of $C_i$ contains $t$
(without variables):  Suppose $\phi$ is satisfiable and for each
component $C_i$ there is at least one tuple $w_i\in C_i$ that does not
contain $t$. Then there is a world-tuple $w \in C_1 \times\cdots\times
C_m$ such that tuple $t$ does not occur in $w$. If there is a mapping
$\theta$ that maps some tuple in $w$ to $t$ and for which
$\theta(\phi)$ is true, then there is also a mapping $\theta'$ such
that $\theta'(w)$ does not contain $t$ but $\theta'(\phi)$ is true.
Thus $t$ is not certain.\punto
\end{proof}
\end{theorem}

As shown in Table~\ref{tab:decision-overview}, tuple certainty is
coNP-complete for c-tables, as it requires to check tautology of local
conditions, which can be arbitrary Boolean formulas.

\begin{theorem}
  \label{th:i-cert}
  Instance certainty is in PTIME for gWSDs.
  \begin{proof}\em
    Given an instance $I$ and a gWSD $W$ representing a relation $R$,
    the problem is equivalent to checking for each world ${\cal A}\in
    \mathit{rep}(W)$ whether (1) $I \subseteq R^{\cal A}$ and (2)
    $R^{\cal A} \subseteq I$. Test (1) is reducible to checking
    whether each tuple from $I$ is certain in $R$, and is thus in
    PTIME (cf. Theorem~\ref{th:t-poss}). For (2), we check in PTIME
    whether there is a tuple different from $t_{\bot}$ in some
    world of $\mathit{rep}(W)$ that is not in the instance $I$. If $W$
    has variables then it cannot represent certain instances.\punto
  \end{proof}
\end{theorem}

\subsection{Tuple and instance q-certainty}

\begin{theorem}
        \label{th:t-qcert}
        Tuple and instance q-certainty are in coNP for gWSDs and
        relational algebra and coNP-hard for WSDs and positive
        relational algebra.
\begin{proof}\em
  Tuple and instance q-certainty are in coNP for gWSDs and full
  relational algebra because gWSDs can be translated polynomially into
  c-tables (see Lemma~\ref{lem:wsdx-c}) and tuple and instance
  q-certainty are in coNP for c-tables and full relational
  algebra~\cite{AKG1991}.
  
  We show coNP-hardness for WSDs and positive relational algebra by a
  reduction from 3DNF-tautology~\cite{gj79}. Given a set \textbf{Y} of
  propositional variables and a set of clauses $c_i = c_{i,1}\wedge
  c_{i,2}\wedge c_{i,3}$ such that for each $i,k$, $c_{i,k}$ is $x$ or
  $\neg x$ for some $x\in\mathbf{Y}$, the 3DNF-tautology problem is to
  decide whether $\bigvee_i c_i $ is true for each truth assignment of
  \textbf{Y}.

  \textbf{Construction.} We create a WSD ${\cal
    W}=(C_1,\ldots,C_{|\textbf{Y}|})$ representing worlds of a
  relation $R$ over schema $R(C, P)$ as follows. For each variable
  $x_i$ in \textbf{Y} we create a component $C_i$ with two local
  worlds, one for $x_i$ and the other for $\neg x_i$. For each literal
  $c_{i,k}$ we create an $R$-tuple $(i,k)$ with id $\tid_{i,k}$. In
  case $c_{i,k} = x_j$ or $c_{i,k} = \neg x_j$, then the schema of
  $C_j$ contains the attributes $R.\tid_{i,k}.C$ and $R.\tid_{i,k}.P$,
  and the local world for $x_j$ or $\neg x_j$, respectively, contains
  the values $(i,k)$ for these attributes. All component fields that
  remained unfilled are finally filled in with $\bot$-values.

  The problem of deciding whether $\bigvee_i c_i$ is a tautology is
  equivalent to deciding whether the nullary tuple $\langle\rangle$ is
  certain in the answer to the fixed positive relational algebra query
  $Q := \pi_{\emptyset}(\sigma_{\phi}((R\ r_1) \times (R\ r_2) \times
  (R\ r_3)))$, where
  \begin{align*}
    \phi :=  (r_1.C = r_2.C \mbox{ and } r_1.C = r_3.C \mbox{ and } r_1.P =1 \mbox{ and } r_2.P = 2 \mbox{ and }  r_3.P = 3).
  \end{align*}

  \textbf{Correctness.}  We prove the correctness of the reduction,
  that is, we show that $\bigvee_i c_i$ is a tautology exactly when
  $\forall {\cal A}\in \mathit{rep}({\cal W}): \langle\rangle \in
  Q^{\cal A}$.
  
  First, assume there is a truth assignment $\nu$ of \textbf{Y} that
  proves $\bigvee_i c_i$ is not a tautology. Then, there exists a
  choice of local worlds of ${\cal W}$ such that for each variable
  $x_i\in\mathbf{Y}$ if $\nu(x_i)=\mathit{true}$ then we choose the
  first local world of $C_i$ and if $\nu(x_i)=\mathit{false}$ then we
  choose the second local world of $C_i$.  Let $w_i$ be the choice for
  $C_i$. Then, ${\cal W}$ defines a world ${\cal A}
  =\mbox{inline}^{-1}(w_1\times\ldots\times w_{|\textbf{Y}|})$ and
  $R^{\cal A}$ contains those tuples defined in the chosen local
  worlds. If $\nu$ proves $\bigvee_i c_i$ is not a tautology, then
  $\nu(\bigvee_i c_i)$ is \textit{false} and, because $\bigvee_i c_i$
  is a disjunction, no clause $c_i$ is \textit{true}. Thus $R^{\cal
    A}$ does not contain tuples $(i,1)$, $(i,2)$, and $(i,3)$ for each
  clause $c_i$. This means that the condition of $Q$ cannot be
  satisfied and thus the answer of $Q$ is empty. Thus the tuple
  $\langle\rangle$ is not certain in the answer to $Q$.
        
  Now, assume there exists a world ${\cal A}\in\mathit{rep}({\cal W})$
  such that $\langle\rangle\not\in Q^{\cal A}$. Then, $R^{\cal A}$
  contains no the set of three tuples $(i,1)$, $(i,2)$, and $(i,3)$
  for any clause $c_i$, because such a triple satisfies the selection
  condition. This means that the choice of local worlds of the
  components in ${\cal W}$ correspond to a valuation $\nu$ that does
  not map all $c_{i,1}$, $c_{i,2}$, and $c_{i,3}$ to true, for any
  clause $c_i$.  Thus $\bigvee_i c_i$ is not a tautology.

  Because by construction $Q^{\cal A}$ is either $\{\}$ or
  $\{\langle\rangle\}$ for any world ${\cal A}\in\mathit{rep}({\cal
  W})$, the same proof also works for instance q-certainty with
  instance $\{\langle\rangle\}$. \punto
\end{proof}
\end{theorem}

\begin{figure}[t!]
\begin{center}
  \begin{align*}
    \mbox{3DNF clause set: } \{c_1 = x_1\wedge x_2\wedge x_3,\hspace*{2em}
    c_2 = x_1\wedge \neg x_2\wedge x_4, \hspace*{2em}
    c_3 = \neg x_1\wedge x_2\wedge \neg x_4\}
  \end{align*}
  \hspace{0.4cm}
  \begin{tabular}{c|ccc}
    $C_1$ & $R.\tid_{1,1}$.(C,P) & $R.\tid_{2,1}$.(C,P) & $R.\tid_{3,2}$.(C,P) \\
    \hline
    $(x_1)$      & (1,1) & (2,1) & $\bot$ \\
    $(\neg x_1)$ & $\bot$ & $\bot$ & (3,2) \\
  \end{tabular}
  \hspace{0.4cm}
  \begin{tabular}{c|c}
    $C_3$ & $R.\tid_{1,3}$.(C,P) \\
    \hline
    $(x_3)$      & (1,3) \\
    $(\neg x_3)$ & $\bot$  \\
  \end{tabular}%
  \hspace*{1.81cm}
  \vspace*{1em}

  \begin{tabular}{c|ccc}
    $C_2$ & $R.\tid_{1,2}$.(C,P) & $R.\tid_{2,2}$.(C,P) & $R.\tid_{3,1}$.(C,P) \\
    \hline
    $(x_2)$      & (1,2) & $\bot$  & (3,1) \\
    $(\neg x_2)$ & $\bot$ & (2,2) & $\bot$  \\
  \end{tabular}
  \hspace{0.4cm}
  \begin{tabular}{c|cc}
    $C_4$ & $R.\tid_{2,3}$.(C,P) & $R.\tid_{3,3}$.(C,P) \\
    \hline
    $(x_4)$      & (2,3) & $\bot$ \\
    $(\neg x_4)$ & $\bot$ & (3,3)  \\
  \end{tabular}
\end{center}
\caption{3DNF clause set encoded as WSD.}
\label{fig:tuple-qcert}
\end{figure}

\begin{example} 
  Figure~\ref{fig:tuple-qcert} gives a 3DNF clause set and its WSD
  encoding. Checking tautology of $H := c_1\vee c_2\vee c_3$ is
  equivalent to checking whether the nullary tuple is certain in the
  answer to the query from the proof of Theorem~\ref{th:t-qcert}.
  Formula $H$ is not a tautology because it becomes \textit{false}
  under the truth assignment $\{x_1\mapsto \mathit{true}, x_2\mapsto
  \mathit{true}, x_3\mapsto \mathit{false}, x_4\mapsto
  \mathit{true}\}$. This is equivalent to checking whether the nullary
  tuple is in the answer to our query in the world ${\cal A}$ defined
  by the first local world of $C_1$ (encoding $x_1\mapsto
  \mathit{true}$), the first local world of $C_2$ (encoding
  $x_2\mapsto \mathit{true}$), the second local world of $C_3$
  (encoding $x_3\mapsto \mathit{false}$), and the first local world of
  $C_4$ (encoding $x_4\mapsto \mathit{true}$). The relation $R^{\cal
    A}$ is $\{\tuple{C:1,P:1}, \tuple{C:2,P:1}, \tuple{C:1,P:2},
  \tuple{C:3,P:1}, \tuple{C:2,P:3}\}$ and the query answer is
  empty.\punto
\end{example}

\vspace{-3mm}

\section{Optimizing gWSDs}
\label{sec:min}

In this section we study the problem of optimizing a given gWSD by
further decomposing its components using the product operation. We
note that product decomposition corresponds to the new notion of
\textit{relational factorization}.  We define this notion and
study some of its properties, like uniqueness and primality or
minimality in the context of relations without variables and the
special $\bot$ symbol. It turns out that any relation admits a unique
minimal factorization, and there is an algorithm, called
\textsf{prime-factorization}, that can compute it efficiently.
We then discuss decompositions of (g)WSD components in the presence of
variables and the $\bot$ symbol.


\subsection{Prime Factorizations of Relations}

\begin{definition}\em
  Let there be schemata $R[U]$ and $Q[U']$ such that $\emptyset\subset
  U'\subseteq U$. A \textit{factor} of a relation $R$ over schema
  $R[U]$ is a relation $Q$ over schema $Q[U']$ such that there exists
  a relation $R'$ with $R = Q\times R'$.
\end{definition}
A factor $Q$ of $R$ is called \textit{proper}, if $Q\not=R$. A factor
$Q$ is \textit{prime}, if it has no proper factors.
%
%
Two relations over the same schema are \textit{coprime}, if they have
no common factors.

\begin{definition}\em
  Let $R$ be a relation. A \textit{factorization} of $R$ is a set
  $\{C_1,\ldots, C_n\}$ of factors of $R$ such that $R =
  C_1\times\ldots\times C_n$.
\end{definition}
In case the factors $C_1,\ldots, C_n$ are prime, the factorization is
said to be \textit{prime}. From the definition of relational product
and factorization, it follows that the schemata of the factors
$C_1,\ldots, C_n$ are a disjoint partition of the schema of $R$.

\begin{proposition}\label{prop:maximal}
  For each relation a prime factorization exists and is unique.
  \begin{proof}\em
    Consider any relation $R$. Existence is clear because $R$ admits
    the factorization $\{R\}$, which is prime in case $R$ is prime.
    
    Uniqueness is next shown by contradiction. Assume $R$ admits two
    different prime factorizations $\{ \pi_{U_1}(R), \dots,
    \pi_{U_m}(R) \}$ and $\{ \pi_{V_1}(R), \dots, \pi_{V_m}(R) \}$.
    Since the two factorizations are different, there are two sets
    $U_i, V_j$ such that $U_i \neq V_j$ and $U_i \cap V_j \neq
    \emptyset$.  But then, as of course $R = \pi_{U - V_j}(R) \times
    \pi_{V_j}(R)$, we have $ \pi_{U_i}(R) = \pi_{U_i}\big(\pi_{U -
      V_j}(R) \times \pi_{V_j}(R)\big) = \pi_{U_i-V_j}(R) \times
    \pi_{U_i \cap V_j}(R)$. It follows that $ \{ \pi_{U_1}(R), \dots,
    \pi_{U_{i-1}}(R), \pi_{U_i - V_j}(R), \pi_{U_i \cap V_j}(R),
    \pi_{U_{i+1}}(R), \dots, \pi_{U_m}(R) \} $ is a factorization of
    $R$, and the initial factorizations cannot be prime.
    Contradiction. \punto
\end{proof}
\end{proposition}

\subsection{Computing Prime Factorizations}
\label{sec:p-decomposition}

This section first gives two important properties of relational
factors and factorizations. Based on them, it further devises an
efficient yet simple algorithm for computing prime factorizations.

\begin{proposition}\label{proposition:decomp-propagation}
  Let there be two relations $S$ and $F$, an attribute $A$ of $S$ and
  not of $F$, and a value $v\in\pi_A(S)$. Then, for some relations
  $R$, $E$, and $I$ holds
  \begin{align*}
    S = F\times R \Leftrightarrow \sigma_{A=v}(S) = F\times E
    \mbox{ and } \sigma_{A\not=v}(S) = F\times I.
  \end{align*}
\begin{proof}\em
 Note that the schemata of $F$ and $R$ represent a disjoint partition
 of the schema of $S$ and thus $A$ is an attribute of $R$.
  
 $\Rightarrow$. Relation $F$ is a factor of $\sigma_{A=v}(S)$ because
 \begin{align*}
   \sigma_{A=v}(S) = \sigma_{A=v}(F\times R) = F\times
   \sigma_{A=v}(R).
 \end{align*}
 Analogously, $F$ is a factor of $\sigma_{A\not=v}(S)$.
  
 $\Leftarrow$. Relation $F$ is a factor of $S$ because
 \begin{align*}
   S = \sigma_{A=v}(S)\cup\sigma_{A\not=v}(S) = F\times E\cup F\times
   I = F\times (E\cup I). \hspace*{9em}\Box
 \end{align*}
\end{proof}
\end{proposition}  

\begin{corollary}\label{corollary:prime}
  A relation $S$ is prime iff $\sigma_{A=v}(S)$ and
  $\sigma_{A\not=v}(S)$ are coprime.
\end{corollary}

%

\medskip\medskip

\begin{figure}[t!]
  \centering

  \framebox[\textwidth]{
    \parbox{12cm}{
      \textbf{algorithm} \textsf{prime-factorization ($S$)}\\
      // Input:  Relation $S$ over schema $S[U]$.\\
      // Result: Prime factorization of $S$ as a set $\mathit{Fs}$ of its prime factors.\\
      
      1. \hspace*{1em} $\mathit{Fs} := \{\{\pi_B(S)\}\mid B\in U, |\pi_B(S)| = 1\}$; $S := S \div \prod\limits_{F \in Fs}{(F)}$;\\
      2. \hspace*{1em} \textbf{if}  $S = \emptyset$ \textbf{then return} $\mathit{Fs}$;\\
      3. \hspace*{1em} \textbf{choose any} $A\in\textsf{sch}(S), v\in\pi_A(S)$ \textbf{such that} $|\sigma_{A=v}(S)|\leq|\sigma_{A\not=v}(S)|$;\\
      4. \hspace*{1em} $\mathit{Q} := \sigma_{A=v}(S)$; $\mathit{R} := \sigma_{A\not=v}(S)$;\\
      5. \hspace*{1em} \textbf{foreach} $F\in \textsf{prime-factorization}(Q)$ \textbf{do}\\
      6. \hspace*{2em} \textbf{if} $(R\div F)\times F = R$ \textbf{then} $\mathit{Fs} := \mathit{Fs}\cup\{F\}$;\\
      7. \hspace*{1em} \textbf{if} $\prod\limits_{F \in Fs}{(F)}\not=S$ \textbf{then} $\mathit{Fs} := \mathit{Fs}\cup\{S\div\prod\limits_{F \in Fs}{(F)}\}$;\\
      8. \hspace*{1em} \textbf{return} $\mathit{Fs}$; }

}
\caption{Computing the prime factorization of a relation.}
  \label{fig:prime-factorization}
\end{figure}

The algorithm \textsf{prime-factorization} given in
Figure~\ref{fig:prime-factorization} computes the prime factorization of an
input relation $S$ as follows. It first finds the trivial prime factors with
one attribute and one value (line 1). These factors represent the prime
factorization of $S$, in case the remaining relation is empty (line 2).
Otherwise, the remaining relation is disjointly partitioned in relations $Q$
and $R$ (line 4) using \textit{any} selection with constant $A=v$ such that
$Q$ is smaller than $R$ (line 3). The prime factors of $Q$ are then probed for
factors of $R$ and in the positive case become prime factors of $S$ (lines 5
and 6). This property is ensured by
Proposition~\ref{proposition:decomp-propagation}. The remainder of $Q$ and
$R$, which does not contain factors common to both $Q$ and $R$,
becomes a factor
of $S$ (line 7). According to Corollary~\ref{corollary:prime}, this factor is
also prime.


\begin{example}
We exemplify our prime factorization algorithm using the following
relation $S$ with three prime factors.
  
  \begin{small}
  \begin{center}
  \begin{tabular}{l|lllll}
    S  & A     & B     & C     & D     & E \\\hline
       & $a_1$ & $b_1$ & $c_1$ & $d_1$ & $e_1$\\
       & $a_1$ & $b_1$ & $c_1$ & $d_1$ & $e_2$\\
       & $a_1$ & $b_1$ & $c_1$ & $d_2$ & $e_1$\\
       & $a_1$ & $b_1$ & $c_1$ & $d_2$ & $e_2$\\
       & $a_2$ & $b_1$ & $c_1$ & $d_1$ & $e_1$\\
       & $a_2$ & $b_1$ & $c_1$ & $d_1$ & $e_2$\\
       & $a_2$ & $b_1$ & $c_1$ & $d_2$ & $e_1$\\
       & $a_2$ & $b_1$ & $c_1$ & $d_2$ & $e_2$\\
       & $a_2$ & $b_2$ & $c_2$ & $d_1$ & $e_1$\\
       & $a_2$ & $b_2$ & $c_2$ & $d_1$ & $e_2$\\
       & $a_2$ & $b_2$ & $c_2$ & $d_2$ & $e_1$\\
       & $a_2$ & $b_2$ & $c_2$ & $d_2$ & $e_2$
  \end{tabular}
  \hspace*{3em}
  \begin{tabular}{|ccc|}
    \hline
    A     & B     & C \\\hline
    $a_1$ & $b_1$ & $c_1$  \\
    $a_2$ & $b_1$ & $c_1$  \\
    $a_2$ & $b_2$ & $c_2$ \\
    \hline
  \end{tabular}
  $\times$
  \begin{tabular}{|c|}
    \hline
    D \\
    \hline
    $d_1$\\
    $d_2$\\
    \hline
  \end{tabular}
  $\times$
  \begin{tabular}{|c|}
    \hline
    E \\
    \hline
    $e_1$\\
    $e_2$\\
    \hline
  \end{tabular}
\end{center}
\end{small}
To ease the explanation, we next consider all variables of the
algorithm followed by an exponent $i$, to uniquely identify their values
at recursion depth $i$.

\medskip 

Consider the sequence of selection parameters $(A,a_1), (D,d_1),
(E,e_1)$.

The relation $S^1$ has no factors with one attribute. We next choose
the selection parameters $(A,a_1)$. The partition $Q^1=\sigma_{A=a_1}(S^1)$
and $R^1=\sigma_{A\not=a_1}(S^1)$ is shown below

\begin{small}
      \begin{center}
  \begin{tabular}{l|lllll}
    $Q^1$  & A     & B     & C     & D     & E \\\hline
       & $a_1$ & $b_1$ & $c_1$ & $d_1$ & $e_1$\\
       & $a_1$ & $b_1$ & $c_1$ & $d_1$ & $e_2$\\
       & $a_1$ & $b_1$ & $c_1$ & $d_2$ & $e_1$\\
       & $a_1$ & $b_1$ & $c_1$ & $d_2$ & $e_2$\\
  \end{tabular}\hspace*{2em}%
  \begin{tabular}{l|lllll}
    $R^1$  & A     & B     & C     & D     & E \\\hline
       & $a_2$ & $b_1$ & $c_1$ & $d_1$ & $e_1$\\
       & $a_2$ & $b_1$ & $c_1$ & $d_1$ & $e_2$\\
       & $a_2$ & $b_1$ & $c_1$ & $d_2$ & $e_1$\\
       & $a_2$ & $b_1$ & $c_1$ & $d_2$ & $e_2$\\
       & $a_2$ & $b_2$ & $c_2$ & $d_1$ & $e_1$\\
       & $a_2$ & $b_2$ & $c_2$ & $d_1$ & $e_2$\\
       & $a_2$ & $b_2$ & $c_2$ & $d_2$ & $e_1$\\
       & $a_2$ & $b_2$ & $c_2$ & $d_2$ & $e_2$
  \end{tabular}
\end{center}  
\end{small}
We proceed to depth two with $S^2 = Q^1$.  We initially find the prime
factors with one of the attributes $A$, $B$, and $C$. We further
choose the selection parameters $(D,d_1)$ and obtain $Q^2$ and $R^2$
as follows
  \begin{center}
  \begin{tabular}{l|ll}
    $Q^2$      & D     & E \\\hline
               & $d_1$ & $e_1$\\
               & $d_1$ & $e_2$\\
  \end{tabular}\hspace*{2em}%
  \begin{tabular}{l|ll}
    $R^2$      & D     & E \\\hline
               & $d_2$ & $e_1$\\
               & $d_2$ & $e_2$\\
  \end{tabular}%
\end{center}
We proceed to depth three with $S^3 = Q^2$. We initially find the
prime factor with the attribute $D$. We further choose the selection
parameters $(E,e_1)$ and obtain $Q^3$ and $R^3$ as follows:
  \begin{center}
  \begin{tabular}{l|ll}
    $Q^3$      & E \\\hline
               & $e_1$
  \end{tabular}\hspace*{2em}%
  \begin{tabular}{l|ll}
    $R^3$      & E \\\hline
               & $e_2$
  \end{tabular}%
\end{center}
We proceed to depth four with $S^4 = Q^3$. We find the only prime
factor $\pi_E(Q^3) = Q^3$ with the attribute $E$ and return the set
$\{Q^3\}$.

At depth three, we check whether $Q^3$ is also a factor of $R^3$. It
is not, and we infer that $Q^3\cup R^3$ is a prime factor of $Q^2$
(the other prime factor $\pi_D(Q^2)$ was already detected). We thus
return $\{\pi_D(Q^2),\pi_E(Q^2)\}$.

At depth two, we check the factors of $Q^2$ for being factors of $R^2$
and find that $\pi_E(Q^2)$ is also a factor of $R^2$, whereas
$\pi_D(Q^2)$ is not. The set of prime factors of $Q^1$ is thus
$\{\pi_E(Q^2), \pi_A(Q^1), \pi_B(Q^1), \pi_C(Q^1), \pi_D(Q^1)\}$,
where $\pi_A(Q^1)$, $\pi_B(Q^1)$, and $\pi_C(Q^1)$ were already
detected as factors with one attribute and one value, and
$\pi_D(Q^1)\}$ is the rest of $Q^1$.

At depth one, we find that only $\pi_E(Q^2)$ and $\pi_D(Q^1)\}$ are
also factors of $R^1$. Thus the prime factorization of $S^1$ is
$\{\pi_E(Q^2), \pi_D(Q^1), \pi_{A,B,C}(S^1)\}$. The last factor is
computed in line 7 by dividing $S^1$ to the product of the factors
$\pi_E(Q^2)$ and $\pi_D(Q^1)\}$.
\punto
\end{example}
 
\medskip

\begin{remark}
  It can be easily verified that choosing another sequence of
  selection parameters, e.g., $(D,d_1)$, $(E,e_1)$ and $(A,a_1)$, does
  not change the output of the algorithm.

  Because the prime factorization is unique, the choice of the
  attribute $A$ and value $v$ (line 3) can not influence it. However,
  choosing $A$ and $v$ such that
  $|\sigma_{A=v}(S)|\leq|\sigma_{A\not=v}(S)|$ ensures that with each
  recursion step the input relation to work on gets halved. This
  affects the worst-case complexity of our algorithm.
  
  In general, there is no unique choice of $A$ and $v$ that halve the
  input relation. There are choices that lead to faster factorizations
  by minimizing the number of recursive calls and also the sizes of
  the intermediary relations $Q$. \punto
\end{remark}

\begin{theorem}\label{th:correctness}
  The algorithm of Figure~\ref{fig:prime-factorization} computes the
  prime factorization of any relation.
\end{theorem}
\begin{proof}
The algorithm terminates, because (1) the input size at the recursion
depth $i$ is smaller (at least halved) than at the recursion depth
$i-1$, and (2) the initial input is finite.
    
We first show by complete induction on the recursion depth that the
algorithm is sound, i.e., it occasionally computes the prime
factorization of the input relation.

Consider $d$ the maximal recursion depth. To ease the rest of the
proof, we uniquely identify the values of variables at recursion depth
$i$ ($1\leq i\leq d$) by an exponent $i$.

Base Case. We show that at maximal recursion depth $d$ the algorithm
computes the prime factorization. This factorization corresponds to
the case of a relation $S^d$ with a single tuple (line 2), where each
attribute induces a prime factor (line 1).

Induction Step. We know that $\mathit{Fs}^{i+1}$ represents the prime
factorization of $S^{i+1} = Q^i$ and show that $\mathit{Fs}^i$
represents the prime factorization of $S^i$.

Each factor $F$ of $Q^i$ is first checked for being a factor of $R^i$
(lines 5 and 6). This check uses the definition of relational
division: the product of $F$ and the division of $R^i$ with $F$ must
give back $R^i$. Using
Proposition~\ref{proposition:decomp-propagation}, each factor $F$
common to $R^i$ and $S^i$ is also a factor of $S^i$. Obviously,
because $F$ is prime in $Q^i$, it is also prime in $S^i$.

We next treat the case when the factors common to $Q^i$ and $R^i$ do
not entirely cover $S^i$ (line 7). Let $P$ be the product of all
factors common to $Q^i$ and $R^i$, i.e., $P = \Pi \mathit{Fs}^i$. Then
there exists $Q^i_*$ and $R^i_*$ such that $Q^i = Q^i_*\times P$ and
$R^i = R^i_*\times P$. It follows that $S^i = Q^i\cup R^i = (Q^i_*\cup
R^i_*) \times P$, thus $(Q^i_*\cup R^i_*)$ is a factor of $S^i$.
Because $Q^i_*$ and $R^i_*$ are coprime (otherwise they would have a
common factor), Corollary~\ref{corollary:prime} ensures that their
union $(Q^i_*\cup R^i_*)$ is prime.

This concludes the proof that the algorithm is sound. The completeness
follows from Proposition~\ref{proposition:decomp-propagation}, which
ensures that the factors of $S^i$, which do not have the chosen
attribute $A$, are necessarily factors of both $Q^i$ and $R^i$ at any
recursion depth $i$. Additionally, this holds independently of the
choice of the selection parameters.\punto
\end{proof}

Our relational factorization is a special case of algebraic
factorization of Boolean functions, as used in multilevel logic
synthesis~\cite{bryant87}. In this light, our algorithm can be used to
algebraically factorize disjunctions of conjunctions of literals. A
factorization is then a conjunction of factors, which are disjunctions
of conjunctions of literals. This factorization is only algebraic,
because Boolean identities (e.g., $x\cdot x=x$) do not
make sense in our context and thus are not considered (Note that
Boolean factorization is NP-hard, see e.g.,~\cite{bryant87}).
  

  


The algorithm of Figure~\ref{fig:prime-factorization} computes prime
factorizations in polynomial time and linear space, as stated by the
following theorem.

\begin{theorem}\label{th:complexity-factorization}
  The prime factorization of a relation $S$ with arity $m$ and size
  $n$ is computable in time $O(m\cdot n\cdot\log n)$ and space
  $O(n+m\cdot \log n)$.
\end{theorem}
\begin{proof}
The complexity results consider the input and the temporary relations
available in secondary storage.

The computations in lines 1, 3, 4, and 7 require a constant amount of scans
over $S$. The number of prime factors of a relation is bounded in its arity.
The division test in line 6 can be also implemented as
$$\pi_{\mathit{sch}(P)}(R) = P\mbox{ and
}|P|\cdot|\pi_{\mathit{sch}(R)-\mathit{sch}(P)}(R)| = |R|.$$
(Here
$\textit{sch}$ maps relations to their schemata). This requires to sort
$P$ and $\pi_{\mathit{sch}(P)}(R)$ and to scan $R$ two times and $P$
one time. The size of $P$ is logarithmic in the size of $Q$, whereas
$Q$ and $R$ have sizes linear in the size of $S$.  The recursive call
in line 5 is done on $Q$, whose size is at most a half of the size of
$S$.

The recurrence relation for the time complexity is then (for
sufficiently large constant $a$; $n$ is the size of $S$ and $m$ is the
arity of $S$)
\begin{align*}
  T(n) &= 7n + m\cdot n\cdot\log n + T\Big(\frac{n}{2}\Big) \\
  &\leq T'(n) = a\cdot m\cdot n\cdot\log n + T'\Big(\frac{n}{2}\Big) = 
  a\cdot m\cdot\overset{\lceil\log n\rceil}{\underset{i=1}{\sum}} \frac{n}{2^i}\cdot\log\Big(\frac{n}{2^i}\Big)\\
  &\leq a\cdot m\cdot\overset{\infty}{\underset{i=1}{\sum}} \frac{n}{2^i}\cdot\log\Big(\frac{n}{2^i}\Big)
  = a\cdot m\cdot n\cdot\log n = O(m\cdot n\cdot\log n).\\
\end{align*}
\vspace*{-1em}

Each factor of $S$ requires space logarithmic in the size of $S$. The
sum of the sizes of the relations $Q$ and $R$ is the size of
$S$. Then, the recurrence relation for the space complexity is
($n$ is the size of $S$ and $m$ is the arity of $S$)
\begin{align*}
  S(n) &= n +  m\cdot \log n + S\Big(\frac{n}{2}\Big)
  =\overset{\lceil\log n\rceil}{\underset{i=1}{\sum}} \Big(\frac{n}{2^i}+m\cdot\log\Big(\frac{n}{2^i}\Big)\Big)\\
  &\leq m\cdot\overset{\infty}{\underset{i=1}{\sum}} \Big(\frac{n}{2^i}+m\cdot\log\Big(\frac{n}{2^i}\Big)\Big)
  = O(n+m\cdot \log n).\vspace*{-1em}
\end{align*}
\punto
\end{proof}

  

We can further trade the space used to explicitly store the temporary
relations $Q$, $R$, and the factors for the time needed to recompute
them. For this, the temporary relations computed at any recursion
depth $i$ are defined \textit{intentionally} as queries constructed
using the chosen selection parameters. This leads to a sublinear space
complexity at the expense of an additional logarithmic factor for the
time complexity.

\begin{proposition}\label{proposition:complexity2-factorization}
  The prime factorization of a relation $S$ with arity $m$ and size
  $n$ is computable in time $O(m\cdot n\cdot \log^2 n)$ and space
  $O(m\cdot\log n)$.
\end{proposition}
\begin{proof}
We can improve the space complexity result of Theorem~\ref{th:complexity-factorization}
in the following way. The temporary relations computed at any recursion depth $i$
are defined \textit{intentionally} as queries constructed using their
schema, say $U^i$, and the chosen selection parameters $(A^i,v^i)$.
  
The relation $Q^j$ at recursion depth $j\leq i$ is 
\begin{align*}
  Q^j = \pi_{\mathit{U}^j}(\sigma_{\phi^Q_j}(S)), \hspace*{2em}\phi^Q_j = \underset{1\leq l\leq j}{\bigwedge} (A^l=v^l)
\end{align*}
The relation $R^j$ is defined similarly and their factors additionally
require to only store their schema. Such queries have the size bounded
in the maximal recursion depth, thus in the logarithm of the input
relation size. At each recursion depth, only an attribute-value pair
needs to be stored. Thus the space complexity becomes
($n$ is the size of $S$ and $m$ is the arity of $S$)
\begin{align*}
  S(n,m) &= m\cdot\log n + S\Big(\frac{n}{2},m-1\Big) \leq \overset{\lceil\log n\rceil}{\underset{i=1}{\sum}} m\cdot \log\frac{n}{2^i}
  \leq \overset{\infty}{\underset{i=1}{\sum}}m\cdot\log\frac{n}{2^i} = m\cdot\log n.
\end{align*}

The time complexity increases, however. All temporary relations need
to be recomputed from the original relation $S$ seven times at each
recursion depth. Thus, in contrast to $T(n,m)$ from the proof of
Theorem~\ref{th:complexity-factorization}, the factor $\frac{1}{2^i}$
does not appear in the new formula of $T(n')$.  The new recurrence
function for $T(n')$ (for sufficiently large $a>0$; $n$ is the size of
the initial $S$ and $m$ is the arity of the initial $S$; $n'$ is
initially $n$) is
\begin{align*}
  T(n') &= 7n + m\cdot n\cdot\log n + T\Big(\frac{n'}{2}\Big) \\
  &\leq T'(n') = a\cdot m\cdot n\cdot\log n + T'\Big(\frac{n'}{2}\Big) = 
  \overset{\lceil\log n\rceil}{\underset{i=1}{\sum}} a\cdot m\cdot n\cdot\log n\\
  & = a\cdot m\cdot n\cdot\log^2 n = O(m\cdot n\cdot\log^2 n).
\end{align*}
\punto
\end{proof}

\begin{remark}
  An important property of our algorithm is that it is polynomial in
  both the arity and the size of the input relation $S$. If the arity
  is considered constant, then a trivial prime factorization algorithm
  (yet exponential in the arity of $S$) can be devised as follows:
  First compute the powerset $\mathit{PS}(U)$ over the set $U$ of
  attributes of $S$. Then, test for each set $U'\in\mathit{PS}(U)$
  whether $\pi_{U'}(S)\times\pi_{U-U'}(S)=S$ holds. In the positive
  case, a factorization is found with factors $\pi_{U'}(S)$ and
  $\pi_{U-U'}(S)$, and the same procedure is now applied to these
  factors until all prime factors are found.
 Note that this algorithm requires time exponential in the arity of
 the input relation (due to the powerset construction).
 Additionally, if the arity of the input relation is constant, then
 the question whether a relation $S$ is prime (or factorizable)
 becomes FO-expressible (also supported by the space complexity given
 in Proposition~\ref{proposition:complexity2-factorization}). 
  \punto
\end{remark}






\subsection{Optimization Flavors}

\begin{figure}[t]
\[
\begin{array}{|cc|}
\hline
R.\tid.A & R.\tid.B \\
\hline
a    & b  \\
\bot & \bot \\
\hline
\end{array}
=
\begin{array}{|c|}
\hline
R.\tid.A\\
\hline
a\\
\hline
\end{array}
\times
\begin{array}{|c|}
\hline
R.\tid.B\\
\hline
b\\
\bot\\
\hline
\end{array}
=
\begin{array}{|c|}
\hline
R.\tid.A\\
\hline
a\\
\bot\\
\hline
\end{array}
\times
\begin{array}{|c|}
\hline
R.\tid.B\\
\hline
b\\
\hline
\end{array}
=
\begin{array}{|c|}
\hline
R.\tid.A\\
\hline
a\\
\bot\\
\hline
\end{array}
\times
\begin{array}{|c|}
\hline
R.\tid.B\\
\hline
b\\
\bot\\
\hline
\end{array}
\]
\caption{Non-unique decompositions of attribute-level WSDs with $\bot$ symbols.}
\label{fig:non-unique}

\vspace{-4mm}

\end{figure}

The algorithm for relational prime factorization can be applied to
find maximal decompositions of (g)WSD components, i.e., minimal
representations of (g)WSDs. Differently from the relational case,
however, the presence of the $\bot$ symbol and of variables may lead
to non-uniqueness and even to non-primality of the (g)WSDs factors
produced by our algorithm. As Figure~\ref{fig:non-unique} shows, the
$\bot$ symbol is one reason for non-unique maximal decompositions of
attribute-level WSDs.

Fortunately, the tuple-level WSDs have maximal decompositions that are
unique modulo the representation of $t_{\bot}$-tuples and can be
efficiently computed by a trivial extension of our algorithm with the
tuple-level constraint. Recall that any tuple
$\tuple{A_1:a_1,\ldots,A_n:a_n}$, where at least one $a_i$ is $\bot$,
is a $t_{\bot}$-tuple.

\begin{proposition}
Any tuple-level WSD has a unique maximal decomposition.
\begin{proof}\em
Let ${\cal W}=\{C_1,\ldots,C_n\}$ be a tuple-level WSD over schema
$(R_1[U_1],\ldots$, $R_k[U_k])$, where $U_j =
(A^1_j,\ldots,A^{n_j}_j)$.

{\bf Construction.} We define a translation $f$ that maps each
component $C_i$ of ${\cal W}$ to an ordinary relation $S_{C_i}$ by
compactifying each tuple $t$ of schema $R_j.d.U_j$ defined by $C_i$
into one value $(t)$ of schema $R_j.d.(U_j)$, where $(U_j)$ is an
attribute. We map all $t_{\bot}$-tuples defined by $C_i$, to the
$\bot$ symbol. We can now apply the algorithm
\textsf{prime-factorization}, where the $\bot$ symbol is treated as
constant.

{\bf Correctness.} We show that there is an equivalence modulo our
translation between maximal decompositions of $S_{C_i}$ and of $C_i$.
Let $\{P_1,\ldots, P_l\}$ and $\{P'_1,\ldots, P'_{l'}\}$ be maximal
decompositions of $C_i$ and $S_{C_i}$, respectively. Because of our
tuple-level constraint, each tuple identifier occurs in the schema of
exactly one $P_j$ and $P'_j$. We show that $l=l'$ and $f(P_j)$ is in
$P'_1,\ldots,P'_j$ modulo the representation of $t_{\bot}$-tuples
(which does not change the semantics of ${\cal W}$).

Assume $l'>l$. Then, there exist two identifiers $d_1$ and $d_2$,
whose tuples are defined in different components of $S_{C_i}$ and the
same component of $C_i$. If $d_1$ and $d_2$ have no $\bot$-values,
then we are in the case of ordinary relations and the algorithm would
have found the same decomposition for $C_i$ and $S_{C_i}$. A
$\bot$-value for one of them cannot influence the values for the other
and thus by treating $\bot$ as a constant, our algorithm would have
found again the same decomposition. Contradiction. We thus have $l=l'$
and the tuples $t$ of an identifier $d$ are defined by a component
$P_j$ of $C_i$ iff $f(t)$ is defined by a $P'_j$ of $S_{C_i}$. The
case of $l>l'$ can be shown similarly.\punto
\end{proof}
\end{proposition}

\begin{figure}[t]
\[
\begin{array}{|cc|}
\hline
R.\tid_1.A & R.\tid_2.A \\
\hline
x    & 1  \\
y    & 2 \\
\hline
\end{array}
=
\begin{array}{|c|}
\hline
R.\tid_1.A\\
\hline
x\\
y\\
\hline
\end{array}
\times
\begin{array}{|c|}
\hline
R.\tid_2.A\\
\hline
1\\
2\\
\hline
\end{array}
=
\begin{array}{|c|}
\hline
R.\tid_1.A\\
\hline
x\\
\hline
\end{array}
\times
\begin{array}{|c|}
\hline
R.\tid_2.A\\
\hline
1\\
2\\
\hline
\end{array}
=
\begin{array}{|c|}
\hline
R.\tid_1.A\\
\hline
y\\
\hline
\end{array}
\times
\begin{array}{|c|}
\hline
R.\tid_2.A\\
\hline
1\\
2\\
\hline
\end{array}
\]
\caption{Equivalent maximal decompositions of tuple-level gWSDs ($x$ and $y$ are variables, the global condition is \textit{true}).}
\label{fig:decomp-var}

\vspace{-4mm}

\end{figure}

The variables are a source of hardness in finding maximal
decompositions of tuple-level gWSDs. By freezing variables and
considering them constant, the three decompositions given in
Figure~\ref{fig:decomp-var} cannot be found by our algorithm.

The gWSD optimization discussed here is a facet of the more general
problem of finding minimal representations for a given g-tabset or
world-set. To find a minimal representation for a given g-tabset ${\bf
A}$, one has to take into account all possible inlinings for the
g-tables of ${\bf A}$ in g-tabset tables. Recall from
Section~\ref{sec:wsd} that we consider a fixed arbitrary inlining
order of the tuples of the g-tables in ${\bf A}$.  Such an order is
supported by common \textit{identifiers} of tuples from different
worlds, as maintained in virtually all representation systems
\cite{IL1984,AKG1991,Gra1991,trio} and {\em exploited}\/ in
practitioner's representation systems such as
\cite{trio,miller06clean}. We note that when no correspondence between
tuples from different worlds has to be preserved, smaller
representations of the same world-set may be possible.

\smallskip

\noindent
{\bf Acknowledgments}.
The authors were supported in part by DFG project grant KO~3491/1-1. The first
author was supported by the International Max Planck Research School for
Computer Science, Saarbr\"ucken, Germany.


\vspace{-5mm}

\bibliographystyle{abbrv}
\bibliography{bibtex}

\end{document}